\documentclass[journal]{IEEEtran}
\usepackage{graphicx}
\usepackage{graphics} 
\usepackage{epsfig} 
\usepackage{mathptmx} 
\usepackage{times} 
\usepackage{amsmath} 
\usepackage{amssymb}  
\usepackage{enumerate}
\usepackage{multirow}
\usepackage{subfigure}
\usepackage{latexsym}
\usepackage{bm}
\usepackage{dsfont}
\usepackage{multicol}
\usepackage{color}
\usepackage{url}
\usepackage{algorithm}
\usepackage{algorithmicx}
\usepackage[algo2e]{algorithm2e}
\usepackage{arydshln}
\usepackage{mathtools}
\usepackage{comment}
\usepackage{cite}
\usepackage{booktabs}
\newtheorem{remark}{Remark}
\newtheorem{assumption}{Assumption}

\newtheorem{proposition}{Proposition}

\newtheorem{theorem}{Theorem}

{\renewcommand{\arraystretch}{2.0}
	\begin{bmatrix}}%
	{\end{bmatrix}
	\renewcommand{\arraystretch}{1.0}}

\def\build#1_#2^#3{\mathrel{\mathop{\kern0pt#1}\limits_{#2}^{#3}}}%

\def\build#1_#2^#3{\mathrel{\mathop{\kern0pt#1}\limits_{#2}^{#3}}}%

\makeatletter
\renewcommand*\env@matrix[1][*\c@MaxMatrixCols c]{%
	\hskip -\arraycolsep
	\let\@ifnextchar\new@ifnextchar
	\array{#1}}
\makeatother

\ifCLASSINFOpdf
\else
\fi

\begin{document}
%
\title{A Dual-level Model Predictive Control Scheme for Multi-timescale Dynamical Systems--Extended Version}
 
%
\author{Xinglong~Zhang,
~Wei~Jiang,
Shuyou Yu,
Xin~Xu,
~Zhizhong~Li
\thanks{Xinglong Zhang, Wei Jiang, and Xin Xu are with the College of Intelligence Science and Technology, National University of Defense Technology, Changsha 410073, China. email: (zhangxinglong18@nudt.edu.cn, jiangweinudt@gmail.com, xuxin\_mail@263.net)}
\thanks{Xinglong Zhang was with the Dipartimento di Elettronica, Informazione e Bioingegneria, Politecnico di Milano, Milan 20133, Italy.}
 
\thanks{Shuyou Yu is with the State Key Laboratory of Automotive Simulation and
Control and the Department of Control Science and Engineering, Jilin University
at NanLing, Changchun 130025, China (e-mail: shuyou@jlu.edu.cn.}
\thanks{Zhizhong Li is with the State Key Laboratory of Disaster Prevention \& Mitigation of Explosion \& Impact, Army Engineering University, Nanjing, Jiangsu 210007, China e-mail: (lizz0607@163.com).}
%
\thanks{This work has been submitted to the IEEE for possible publication. Copyright may be transferred without notice, after which this version may no longer be accessible. The work was supported by the National Natural Science Foundation of China under Grant 61825305, 62003361, and the National Key R$\&$D Program of China
	2018YFB1305105.}}
 
\markboth{}
{Shell \MakeLowercase{\textit{et al.}}: A Dual-level Model Predictive Control Scheme for Multi-timescale Dynamical Systems--Extended Version}
 
%


\IEEEtitleabstractindextext{%
\begin{abstract}
\color{black}So far, many {\color{black}control algorithms have been developed for singularly perturbed systems.  However, in many industrial processes, enforcing closed-loop fast-slow dynamics for peculiarly non-separable ones is a prior request and a crucial issue to be resolved. Aiming at the above problem, this paper presents two dual-level model predictive control (MPC) algorithms for two-timescale dynamical systems with unknown bounded disturbance and input constraint. The proposed algorithms, each one composed of two regulators working in slow and fast time scales, are designed to generate closed-loop separable dynamics at the high and low levels. As a key feature, the proposed algorithms are not only suitable for singularly perturbed systems, but also capable of imposing separable closed-loop performance for dynamics that are non-separable and strongly coupled. The recursive feasibility and convergence properties are proven under suitable assumptions. The simulation results on controlling a Boiler Turbine (BT) system, including the comparisons with other classic controllers are reported, which show the effectiveness of the proposed algorithms.}
\end{abstract}
 
}

\maketitle

\IEEEdisplaynontitleabstractindextext

%
\IEEEpeerreviewmaketitle

\section{Introduction}
\noindent
Many industrial processes are characterized by separable fast-slow dynamics, which can be called ``multi-timescale dynamic systems". In a multi-timescale dynamic system, for a given constant input signal, some of the output variables reach their steady-state values fast while the other ones may experience a longer transient period, see for instance~\cite{naidu2002singular,kokotovic1999singular,mishchenko1994asymptotic}. A widely-acceptable approach for the control of such systems consists in resorting to hierarchical control synthesis that possibly relies on singular perturbation theory (see the book~\cite{naidu1988singular}),  where time-scale separation technique is adopted to define regulators at different control frequencies to guarantee the stability and performance of the dynamics associated with the adopted channels. In addition to singularly perturbed systems, there might be systems whose dynamics are not separable but must be controlled in the same way with a multi-rate control setting, see for instance the control of a Boiler Turbine (BT) system considered in~\cite{RV2008}. {\color{black}In this case, usually, the crucial controlled variables of the considered system must be adjusted at a faster rate to meet the control performance requirement, while other outputs can be controlled more smoothly in a slower time scale.} {\color{black}As another example, the control applications for robots fit the above control problem setting. For instance, in the race car control problem, see~\cite{liniger2015optimization,verschueren2014towards}, the velocity is the major concern to reach the destination with minimum-time periods, while the reference tracking goal is of less importance. In urban autonomous driving applications, see~\cite{paden2016survey}, lane-keeping and precise tracking are the prior concerns while the velocity can be controlled in a smoother manner.}

Model predictive control (MPC) is an advanced process control technique, widely used in industrial processes, such as chemical plants and smart grids, see~\cite{forbes2015model,qin2003survey,li2017nonlinear,jin2018improved,roshany2013kalman},  in robotics, see~\cite{zhou2017real,santoso2018robust}, in urban traffics, see~\cite{zhou2017two}, and in computing data center, see~\cite{fang2017thermal}. In MPC, the control problem is reformulated as an optimization one, that has to be solved on-line iteratively. This allows to explicitly consider the control, state, and output constraints in the control problem. At any generic time instant $k$, a finite horizon optimization problem must be solved to compute the optimal control sequence. Only the first element is applied to the system, and the state and output variables are updated. The optimization problem is then repeated at the next instant $k+1$. 
 
{\color{black}In the framework of MPC,  many solutions have been developed based on time-scale separation technique for systems characterized by open-loop separable dynamics. Among them, in~\cite{chen2012composite} a fast-slow MPC solution has been proposed for control of nonlinear singularly perturbed systems, and the extensions to large-scale systems and to the dynamic optimization of economic cost have been addressed in ~\cite{chen2011model,ellis2013economic}. The algorithms utilize the reduced-order models of the original system, with model couplings between fast and slow scales disregarded, which leads to a decentralized controller design. In~\cite{ma2017slow,miao2015fast}, controllers designed with unitary slow or fast sampling period have been proposed for linear singularly perturbed systems with control saturation. An MPC design with closed-loop property guarantee has been presented in~\cite{wogrin2010mpc} for continuous-time singularly perturbed systems. In~\cite{niu2009two} a decentralized controller design is proposed using input-output models. In the application aspect, notable contributions can be found in,  \cite{brdys2008hierarchical} for integrated wastewater treatment systems, \cite{ohshima1994multirate} for control of a polymerization reactor, \cite{van2009time} for greenhouse climate management, and \cite{sanila2015simultaneous} for control of flexible joint manipulator. As a summary comment, almost all the aforementioned approaches are tailored for systems with clearly different dynamics due to their dependencies on singular perturbation theory, and the closed-loop stability relies on the assumption of the open-loop dynamics being separable. The control performance of such controllers could be hampered for the case that the dynamical motions are open-loop non-separable.}
 
Motivated by the above problems, this work concerns designing dual-level algorithms based on MPC for linear multi-timescale dynamical systems with unknown bounded disturbances and input constraint, to exhibit closed-loop separable dynamic behaviors. 
A novel dual-level MPC (D-MPC) algorithm is initially presented for systems with bounded disturbances. The approach consists of dual levels. At the high level,  an MPC problem is designed  for the control commitment in the slow control channel, while a shrinking horizon MPC is designed in the basic time scale to derive satisfactory closed-loop fast dynamics.  {\color{black}As a core contribution, an incremental form of D-MPC, i.e., \emph{Incremental} D-MPC is proposed to improve the performance associated with the fast (crucial) output and to compensate for possible unknown piece-wise constant disturbances. Under suitable assumptions, the recursive feasibility and convergence properties of the proposed D-MPC and \emph{Incremental} D-MPC are proven.}
Compared with the aforementioned works, see such as~\cite{chen2011model,chen2012composite,wogrin2010mpc,niu2009two,van2009time}, the proposed ones show advantageous points in two aspects. First, the proposed ones utilize consistent models for controller designs at both time scales, being suitable for the ones with strong coupling effects that exhibit non-separable dynamics. Importantly, the controller at each time scale concerns the overall control performance in a cooperative way, which is significantly different from the decentralized independent design in the above methods.
 
{\color{black}A similar problem has been addressed in \cite{zhang2018}, however, the control scheme described in this paper shows a significant improvement for the following reasons: i) The algorithm in  \cite{zhang2018} is proposed for systems described by impulse responses, with a special focus on the viewpoint of application, while this paper presents novel solutions on the theoretical developments based on a state-space formulation, with verified closed-loop recursive feasibility and stability. ii) Due to the usage of impulse response representation, the concerned system in \cite{zhang2018} is assumed to be strictly stable. {\color{black}Whereas in this paper the restriction is relaxed, i.e., the considered model is assumed stabilizable.} iii)  In~\cite{zhang2018},  the input associated with the slow dynamics is only manipulated in the slow time scale. This could lead to control performance degradation especially for systems that are strongly coupled. To solve this problem, we allow the ``slow" control variable to be refined in the fast time scale to improve the control performance.}
 
A two-layer control structure has been proposed in~\cite{picasso2016hierarchical}, but the control problem considered is different. Indeed, it is designated for the coordination of large-scale independent subsystems that must produce a global constant throughput. The approach proposed in this paper commits to enforcing separable closed-loop dynamics for strongly coupled systems and robust design under uncertainties is addressed. Hence, the control framework, and the techniques adopted in this paper are significantly different from that in~\cite{picasso2016hierarchical}.
 
The rest of the paper is organized as follows. Section 2 presents the problem description and the proposed control structure. The
MPC problems at the high and low levels of the D-MPC are introduced in
Section 3, while the \emph{Incremental} D-MPC algorithm is described in Section 4. Simulation example  concerning
the BT control is studied in Section 5, while some conclusions
are drawn in Section 6.  Proofs of the theoretical results are given in the Appendix.\\
\noindent\textbf{Notation:} for a given set of variables $z_{i}\in{\mathbb{R}}^{q_{i}}$,
$i=1,2,\dots,M$, we define $(z_{1}, z_{2}, \cdots, z_{M})=[\,z_{1}^{\top}\ z_{2}^{\top}\ \cdots\ z_{M}^{\top}\,]^{\top}\in{\mathbb{R}}^{q}$, where $q= \sum_{i=1}^{M}q_{i}$.
We use $\mathbb{C}$ to denote the set of the complex plane.
Given a matrix $P$, we use the symbol $P^{\top}$ to denote its transpose. For a generic variable $z$, we denote $\Delta z(k)=z(k)-z(k-1)$, where $k$ is the discrete-time index.
We use $\|x\|_Q^2$ to represent $x^{\top}Qx$. We use $\mathbb{N}$ and $\mathbb{N}_+$ to denote the set of non-negative and positive  integers respectively. Given two sets $A$ and $B$, we denote $A\times B$ as the Cartesian product.
Given the signal $v$, we denote $\overrightarrow{v}(k:k+N-1)$ the sequence $v (k)\ldots v(k+N-1)$, where $N$ is a positive integer.
 
\section{Problem formulation}
The system to be controlled is described by a discrete-time linear system consisting of two interacting subsystems expressed as{\color{black}
\begin{subequations}\label{Eqn:sigma_s_f}
	\begin{align}
	&\Sigma_s:\ \left\{\begin{array}{l}
	x_s(h+1)=A_{ss}x_s(h)+A_{sf}x_f(h)+B_{ss}u_s(h)+\\
\ \ \ \ \ \ \ \ \ \ \ \ \	B_{sf}u_f(h)+d_s(h)\\
	[0.2cm]y_s(h)=C_{ss}x_s(h),
	\end{array} \right.\qquad\label{Eqn:sigma_s}\\
	&\Sigma_f:\ \left\{\begin{array}{l}
	x_f(h+1)=A_{fs}x_s(h)+A_{ff}x_f(h)+B_{fs}u_s(h)+\\
\ \ \ \ \ \ \ \ \ \ \ \ \	B_{ff}u_f(h)+d_f(h)\\
	[0.2cm]y_f(h)=C_{ff}x_f(h),
	\end{array} \right.\qquad\label{Eqn:sigma_f}
	\end{align}
\end{subequations}}
where $u_s\in \mathbb{R}^{m_s}$, $x_s\in \mathbb{R}^{n_s}$, $y_s\in \mathbb{R}^{p_s}$, and $d_s\in\mathcal{D}_s\subseteq\mathbb{R}^{n_s}$ are the input, state, output variables and {\color{black}unmeasured disturbance} belonged to $\Sigma_s$,
while
$u_f\in \mathbb{R}^{m_f}$, $x_f\in \mathbb{R}^{n_f}$, $y_f\in \mathbb{R}^{p_f}$, {\color{black}and $d_f\in\mathcal{D}_f\subseteq\mathbb{R}^{n_f}$}  are the ones  associated with $\Sigma_f$, {\color{black}$\mathcal{D}_s$ and $\mathcal{D}_f$ are compact sets}, $h$ is a basic discrete-time scale index, the matrices $A_{\ast}$, $B_{\ast}$ (where $\ast$ is $sf$ or $fs$ in turn) represent the couplings between $\Sigma_s$ and $\Sigma_f$ through the state and input variables respectively.
 
Similar to~\cite{zhang2018}, in this paper, models \eqref{Eqn:sigma_s} and \eqref{Eqn:sigma_f} are assumed to satisfy at least one of the following scenarios:
\begin{itemize}
	\item $\Sigma_s$ is characterized by a slower dynamics in contrast to $\Sigma_f$ in the sense that some of the triples $(u_f,\, x_f,\, y_f)$ reach their final steady-state values fast while the other
	ones, i.e. $(u_s,\, x_s,\, y_s)$ may have begun their main dynamic motions, see the examples in~\cite{naidu2002singular,kokotovic1999singular,mishchenko1994asymptotic};
	\item even if the dynamics of $\Sigma_s$ and $\Sigma_f$ might not be strictly separable, however they must be controlled in a multi-rate fashion, e.g., the triples $(u_f,\, x_f,\, y_f)$ must react promptly to respond to operation (reference) variations while the triples $(u_s,\, x_s,\, y_s)$ can be controlled in a smoother fashion, see for instance~\cite{RV2008}.
\end{itemize}
{\color{black}Notice that, in a singularly perturbed system, couplings between different time scales are weak, hence the overall model is usually decomposed into decentralized reduced-order models with different time scales.  However, in our case, interactions of the considered system can be non-separable, i.e.,  couplings between subsystems might be strong. To cope with possible coupling effects, system~\eqref{Eqn:sigma_s_f} can be regarded as a whole being later used in the controller design.}
Combining~\eqref{Eqn:sigma_s},~\eqref{Eqn:sigma_f}, the overall system is written as
{\color{black}\begin{equation}\label{Eqn:CL}
\Sigma:\ \left\{\begin{array}{l}
x(h+1)=Ax(h)+Bu(h)+d(h)\\
[0.2cm]y(h)=Cx(h),
\end{array}\right. \qquad
\end{equation}}
where $u=(u_s,\ u_f)\in\mathbb{R}^m$, $m=m_s+m_f$, $x=(x_s,\, x_f)\in\mathbb{R}^n$, $n=n_s+n_f$, $y=(y_s,\, y_f)\in \mathbb{R}^{p}$, $p=p_s+p_f$, {\color{black}the unknown disturbance $d=(d_s,d_f)\in\mathcal{D}_s\times \mathcal{D}_f=\mathcal{D}$.} The diagonal blocks of the collective state transition matrix $A$ and  input matrix $B$ are  $A_{ss}$, $A_{ff}$ and $B_{ss}$, $B_{ff}$ respectively; whereas their non-diagonal blocks correspond to the coupling terms of the state and input variables between $\Sigma_s$ and $\Sigma_f$. The collective output matrix is $C=\text{diag}(C_{ss},\ C_{ff})$. \\
%
The control objectives to be achieved are introduced here.
\begin{itemize}
	\item[(i)] {\color{black}\textbf{Setpoint regulation}:} for a given reference value $y_{r}=(y_{s,r},\,y_{f,r})$, we aim to drive
	\begin{subequations}\label{eq:output_r}	
		\begin{align}	
		y_s(h)\rightarrow y_{s,r},\label{Eqn:y_s-obje}\\
		y_f(h)\rightarrow y_{f,r}\label{Eqn:y_f-obje}
		\end{align}
	\end{subequations}
	
	\item[(ii)] \textbf{Input constraint}:  enforce the input constraint of the type
	\begin{subequations}\label{Eqn:out-in-con}
		\begin{align}
		u_s(h)\in&\mathcal{U}_s,\label{eq:u_s_constraint}\\
		u_f(h)\in& \mathcal{U}_f,\label{eq:u_f_constraint}	
		\end{align}	
	\end{subequations}
	where 
	$\mathcal{U}_s$, $\mathcal{U}_f$ are convex sets, {\color{black} 
		$\mathcal{U}=\mathcal{U}_s\times \mathcal{U}_f$.}
\end{itemize}
	The following assumption is assumed to hold:
	\begin{assumption}\label{assump:A}
		\begin{enumerate}[(1)]	
		{\color{black}		
		 \item The pair $(A,B)$ is stabilizable;
		\item\label{assum1:nozero}
		 $m_f=p_f$. Also, a steady-state pair $\left(u_r, x_r\right)$ exists associated with the reference $y_r$, such that $x_r=Ax_r+Bu_r$, $y_r=Cx_r$, $x_r=(x_{s,r},\, x_{f,r})$, and $u_r=(u_{s,r},\, u_{f,r})\in\mathcal{U}_s\times \mathcal{U}_f$.}
		\end{enumerate}
	\end{assumption}
\begin{remark}
{\color{black}Assumption~\ref{assump:A}.\eqref{assum1:nozero} allows the considered system~\eqref{Eqn:CL} being non-square, i.e., $m_s\neq p_s$. Specifically, for $m\leq p$, given reachable setpoint $y_r$, one can compute $(x_r,u_r)=\Phi^{\dagger}(0,y_r)$, where $$\Phi=\begin{bmatrix}I-A&-B\\C& 0
	\end{bmatrix},$$ $\Phi^{\dagger}=(\Phi^{\top}\Phi)^{-1}\Phi$.	For $p<m$, multiple steady-state solutions might exist associated with $y_r$. In this case, one can select a suitable steady-state pair via solving a static optimization problem with respect to the decision variable $x$, $u$, optimizing a user-specified economic performance index subject to the constraints $x=Ax+Bu$, $y_r=Cx$, 
	$u\in\mathcal{U}$.} 
\end{remark}
	

In principle, a centralized MPC problem with respect to $\Sigma$ can be solved  so as to achieve the above objectives. However, the resulting control performance might be hampered in the aforementioned scenarios due to the conflicting requirements of the sampling period and prediction horizon for $\Sigma_s$ and $\Sigma_f$ respectively.
 
\begin{figure}[h!]
	\center
	\includegraphics[width=0.6\columnwidth]{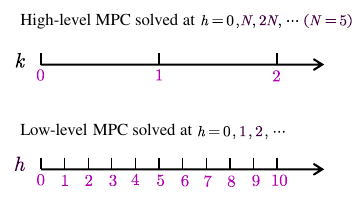}
	\caption{{\color{black}The time indices adopted in different levels:  $h=1,2,\cdots$ denotes the basic (fast) time instant, while $h=0,N,2N,\cdots$ denotes the slow time instant in the basic time scale. The above figure shows a special case with $N=5$.}
	}
	\label{fig:timeaxis}
\end{figure}

For this reason, a dual-level MPC (D-MPC) 
is initially proposed in this paper to fulfill the aforementioned control objectives.
As shown in Figure~\ref{fig:timeaxis}, at the high level, a slow time scale $k$ associated with $N\in\mathbb{N}$ period of the basic time scale $h$ is adopted to define an MPC problem concerning the sampled version of $\Sigma$. 
The computed values of the control actions at this level, $u_f^{\scriptscriptstyle[N]}(k)$, $u_s^{\scriptscriptstyle[N]}(k)$,, are held constant  within the long sampling time interval $[kN,kN+N)$, i.e., $\bar u_f(h)=u_f^{\scriptscriptstyle[N]}(k)$, $\bar u_s(h)=u_s^{\scriptscriptstyle[N]}(k)$ for all $h\in [kN,kN+N)$. At the low level, a shrinking horizon MPC is designed at the basic time scale to refine control actions with additional corrections (i.e., $\delta u_f(h)$, $\delta u_s(h)$) in order to derive satisfactory short-term transient associated with the closed-loop fast dynamics and to account for possible disturbances.
 
The overall control actions of the D-MPC regulator are described by
\begin{subequations}\label{eq:input_contributions}
	\begin{align}
	u_s(h)&=\bar u_s(h)+\delta u_s(h),\label{Eqn:slow-in}\\
	u_f(h)&=\bar u_f(h)+\delta u_f(h)\label{Eqn:fast-in}
	\end{align}
\end{subequations}
where
\begin{itemize}
	\item  the control actions $\bar u_s(h)$ and $\bar u_f(h)$ will be computed by solving an MPC problem according to receding horizon principle in the slow time scale to fulfill objective \eqref{Eqn:y_s-obje} and \eqref{eq:u_s_constraint}, \eqref{eq:u_f_constraint};
	\item the corrections $\delta u_s(h)$ and $\delta u_f(h)$ will be defined by a shrinking horizon MPC regulator running in the basic time scale to fulfill objective \eqref{Eqn:y_f-obje} and to enforce constraints \eqref{eq:u_s_constraint}, \eqref{eq:u_f_constraint}.
\end{itemize}
%
Moreover, to further improve the control behavior of the fast controlled variables, an \emph{Incremental} D-MPC algorithm is also proposed (see Section~\ref{sec:iD-MPC}). To be specific, this version includes integral actions at the two levels and, a prior explicit design of the output trajectory of $y_f$ at the slow time scale to enforce  $y_f$ to the reference value or its neighbor promptly. 
{\color{black}A brief diagram of the proposed approaches is displayed in Figure~\ref{fig:control_scheme}.
\begin{figure}[h!]
	\center
	\includegraphics[width=0.8\columnwidth]{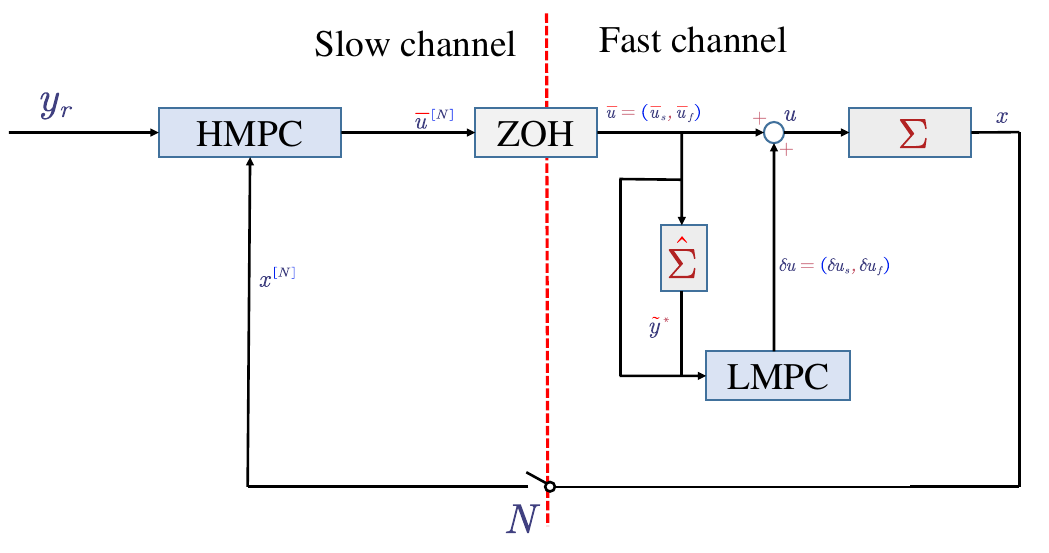}
	\caption{{\color{black}A brief diagram of the proposed control scheme: HMPC (LMPC) stands for the MPC at the higher (lower) level, while ZOH is the zero order holder.}}
	\label{fig:control_scheme}
\end{figure}
}
 
\section{D-MPC algorithm}
In this section, the D-MPC algorithm consisting of an  MPC at the high level and a shrinking horizon MPC at the low level is devised. 
\subsection{MPC at the high level}
In order to design the high-level regulator in the slow time scale $k$ {\color{black}(see again Figure~\ref{fig:timeaxis})}, 
denote by $u_{\ast}^{\scriptscriptstyle[N]}$, $x_{\ast}^{\scriptscriptstyle[N]}$, $y_{\ast}^{\scriptscriptstyle[N]}$, and $d_{\ast}^{\scriptscriptstyle[N]}$  the samplings of $u_{\ast}$, $x_{\ast}$, $y_{\ast}$, and $d_{\ast}$  (where $\ast$ is $s$ or $f$, in turn) and by $u^{\scriptscriptstyle[N]}=(u_s^{\scriptscriptstyle[N]},\,u_f^{\scriptscriptstyle[N]})$, $x^{\scriptscriptstyle[N]}=(x_s^{\scriptscriptstyle[N]},\,x_f^{\scriptscriptstyle[N]})$, $y^{\scriptscriptstyle[N]}=(y_s^{\scriptscriptstyle[N]},\,y_f^{\scriptscriptstyle[N]})$, {\color{black}and $d^{\scriptscriptstyle[N]}=(d_s^{\scriptscriptstyle[N]},\,d_f^{\scriptscriptstyle[N]})$} the samplings of the input, state, and output variables corresponding to the time scale $k$.  Hence, the sampled version of \eqref{Eqn:CL} with $N$ period is given as
{\color{black}
\begin{equation}\label{Eqn:CL_k}
\Sigma^{\scriptscriptstyle[N]}:\ \left\{\begin{array}{l}
x^{\scriptscriptstyle[N]}(k+1)=A^{\scriptscriptstyle[N]}x^{\scriptscriptstyle[N]}(k)+B^{\scriptscriptstyle[N]}u^{\scriptscriptstyle[N]}(k)+\tilde d^{\scriptscriptstyle[N]}(k)\\
[0.2cm]y^{\scriptscriptstyle[N]}(k)=Cx^{\scriptscriptstyle[N]}(k),
\end{array}\right. \qquad
\end{equation}
}
where $A^{\scriptscriptstyle[N]}=A^{N}$, $B^{\scriptscriptstyle[N]}=\sum_{j=0}^{N-1}A^{N-j-1}B$, {\color{black}$\tilde d^{\scriptscriptstyle[N]}(k)=f^{\scriptscriptstyle[N]}_{\delta u}(k)+d^{\scriptscriptstyle[N]}(k)$,  $d^{\scriptscriptstyle[N]}(k)=\sum_{i=0}^{N-1}A^id(kN+i)$, $f^{\scriptscriptstyle[N]}_{\delta u}(k)=\sum_{i=0}^{N-1}A^iB\delta u(kN+i)$ is due to the control at the low level.}
{\color{black}
 
The following proposition can be stated for $\Sigma^{[N]}$:
\begin{proposition}\label{prop:obser-con}
	The pair $(A^{\scriptscriptstyle[N]},\,C)$ is detectable if $(A,\,C)$ is detectable.
\end{proposition}
 
Also, the following assumption about $\Sigma^{[N]}$ is assumed to be holding:
\begin{assumption}\label{assum:stab-0}
	The pair $(A^{\scriptscriptstyle[N]},\,B^{\scriptscriptstyle[N]})$ is stabilizable.
\end{assumption}
}
{\color{black}
\begin{remark}
Note that, slightly different from the detectability condition in Proposition~\ref{prop:obser-con}, to meet the stabilizability requirement of $\Sigma^{\scriptscriptstyle[N]}$, we require Assumption~\ref{assum:stab-0} to be verified a posteriori once the sampling period $N$ is chosen. This is due to the fact that, starting from the stabilizability of $(A,\,B)$, there is no guarantee the resultant sampled pair $(A^{\scriptscriptstyle[N]},\,B^{\scriptscriptstyle[N]})$ is also stabilizable. A simple example to illustrate this point is as follows: consider a single-input single-output system described by $x(h+1)=-x(h)+u(h)$, which is obviously stabilizable. However, the $N=2$ period sampled version $x^{\scriptscriptstyle[2]}(k+1)=(-1)^2x^{\scriptscriptstyle[2]}(k)+(-1+1)u^{\scriptscriptstyle[2]}(k)=x^{\scriptscriptstyle[2]}(k)$ is not stabilizable.
\end{remark}
}
%
{\color{black}
As the disturbance term $\tilde d^{\scriptscriptstyle[N]}$ is unknown, for prediction, we introduce the following nominal model:
\begin{equation}\label{Eqn:CL_k_nominal}
\hat \Sigma^{\scriptscriptstyle[N]}:\ \left\{\begin{array}{l}
\hat x^{\scriptscriptstyle[N]}(k+1)=A^{\scriptscriptstyle[N]}\hat x^{\scriptscriptstyle[N]}(k)+B^{\scriptscriptstyle[N]}u^{\scriptscriptstyle[N]}(k)\\
[0.2cm]\hat y^{\scriptscriptstyle[N]}(k)=C\hat x^{\scriptscriptstyle[N]}(k),
\end{array}\right. \qquad
\end{equation}
Notice that, if $\hat x(kN)=\hat x^{\scriptscriptstyle[N]}(k)$ and the control $u(h)=u^{\scriptscriptstyle[N]}(k),\ \forall h\in[kN,kN+N)$, it holds that   $\hat x(kN+N)=\hat x^{\scriptscriptstyle[N]}(k+1)$ and $\hat y(kN+N)=\hat y^{\scriptscriptstyle[N]}(k+1)$.\\
}
With \eqref{Eqn:CL_k_nominal}, it is now possible to state the MPC problem at the high level. 
At each slow time-step $k$ we solve an optimization problem according to receding horizon principle as follows:
\begin{equation}
\begin{array}{cc}
\min & {J_{{\rm\scriptscriptstyle H}}}\\
{\scriptstyle \overrightarrow{u^{\scriptscriptstyle[N]}}{(k:k+N_{\rm\scriptscriptstyle H}-1)}}\\
\end{array}\label{Eqn:HLoptimiz}
\end{equation}
where
\begin{equation}
\begin{array}{cl}
J_{{\rm\scriptscriptstyle H}}=&\sum_{i=0}^{N_{\rm\scriptscriptstyle H}-1}\big(\|\hat y^{\scriptscriptstyle[N]}(k+i)-y_r\|_{Q_{\rm\scriptscriptstyle H}}^{2}+\|u^{\scriptscriptstyle[N]}(k+i)-u_r\|_{R_{\rm\scriptscriptstyle H}}^{2}\big)\\
& +\|\hat x^{\scriptscriptstyle[N]}(k+N_{\rm\scriptscriptstyle H})-x_r\|_{P_{\rm\scriptscriptstyle H}}^{2},
\end{array}\label{Eqn:HL_cost}
\end{equation}	
and where $N_{\rm\scriptscriptstyle H}>0$ is the adopted prediction horizon.
The parameters ${Q_{\rm\scriptscriptstyle H}}\in \mathbb{R}^{p\times p}$, ${R_{\rm\scriptscriptstyle H}}\in \mathbb{R}^{m\times m}$ are positive definite and symmetric weighting matrices, while  ${P_{\rm\scriptscriptstyle H}}\in \mathbb{R}^{n\times n}$ is computed as the solution to the Lyapunov equation described by
\begin{equation}\label{Eqn:lyap}
F_{\rm\scriptscriptstyle H}^{\top} P_{\rm\scriptscriptstyle H} F_{\rm\scriptscriptstyle H}-P_{\rm\scriptscriptstyle H}=-(C^{\top}Q_{\rm\scriptscriptstyle H}C+K_{\rm\scriptscriptstyle H}^{\top}R_{\rm\scriptscriptstyle H} K_{\rm\scriptscriptstyle H})
\end{equation}
where matrix $F_{\rm\scriptscriptstyle H}=A^{\scriptscriptstyle[N]}+B^{\scriptscriptstyle[N]}K_{\rm\scriptscriptstyle H}$ is Schur stable and $K_{\rm\scriptscriptstyle H}$ is a stabilizing gain matrix.\\
The optimization problem \eqref{Eqn:HLoptimiz} is performed under the following constraints:
\begin{enumerate}[(1)]
	\item  the dynamics \eqref{Eqn:CL_k_nominal};
	{\color{black}
	\item the input constraint
	\begin{equation*}
	\begin{array}{ccc}
	u^{\scriptscriptstyle[N]}(k+i)\in \bar{\mathcal{U}},
	\end{array}
	\end{equation*}
	where $ \bar{\mathcal{U}}$  is a tightened convex set of $\mathcal{U}$, i.e., $  \bar{\mathcal{U}}\subseteq\mathcal{U}$;
	\item the terminal state constraint
	\begin{equation*}
	\hat x^{\scriptscriptstyle[N]}(k+N_{\rm\scriptscriptstyle H})\in \mathcal{X}_{\rm\scriptscriptstyle F}^s,
	\end{equation*}}
\end{enumerate}
	{\color{black}where  $\mathcal{X}_{\rm\scriptscriptstyle F}^s\subseteq\mathcal{X}_{\rm\scriptscriptstyle F}$,} the set $\mathcal{X}_{\rm\scriptscriptstyle F}$ is chosen as a positively invariant set for system~\eqref{Eqn:CL_k_nominal} controlled with the stabilizing control law $u^{\scriptscriptstyle[N]}(k)=K_{\rm\scriptscriptstyle H} (\hat x^{\scriptscriptstyle[N]}(k)-x_r)+u_r$, satisfying $K_{\rm\scriptscriptstyle H}(\mathcal{X}_{\rm\scriptscriptstyle F}\ominus x_r) \subseteq \mathcal{U}\ominus u_r$. 	{\color{black}To guarantee the recursive feasibility under uncertainties, ${\mathcal{X}}_{\rm\scriptscriptstyle F}^s$ is selected such that: for any $z(k)\in {\mathcal{X}}_{\rm\scriptscriptstyle F}$, the successive state under the above stabilizing control law satisfies  $z(k+1)\in {\mathcal{X}}_{\rm\scriptscriptstyle F}^s$.
Let $\overrightarrow{u^{\scriptscriptstyle[N]}}{(k:k+N_{\rm\scriptscriptstyle H}-1|k)}$ be the optimal solution to optimization \eqref{Eqn:HLoptimiz}. Only the first element $u^{\scriptscriptstyle[N]}(k|k)$ is used for designing the regulator at the low level that will be applied in the fast time scale. At the next time $k+1$, the optimization is repeated according to receding horizon principle.}

\subsection{Shrinking horizon MPC at the low level}
{\color{black}Assume now to be at a specific basic time instant $h=kN$ (correspond to the slow time instant $k$, see Figure~\ref{fig:timeaxis})} such that the high-level optimization problem \eqref{Eqn:HLoptimiz} with cost~\eqref{Eqn:HL_cost} has been successfully solved. Thus the computed values of the input $u^{\scriptscriptstyle[N]}(k)=(u_s^{\scriptscriptstyle[N]}(k),\,u_f^{\scriptscriptstyle[N]}(k))$ and the one-step ahead state prediction $\hat x^{\scriptscriptstyle[N]}(k+1|k)$ are available. Let us focus on the output performance in the fast time scale within the interval $h\in[kN,\,kN+N)$.
Denoting by  $\tilde{y}(h)=(\tilde y_s(h),\,\tilde y_f(h))$ the output resulting from~\eqref{Eqn:CL_k_nominal} with $u(h)=u^{\scriptscriptstyle[N]}(\lfloor{h/N}\rfloor)$, the component $\tilde y_f(h)$ may expect undesired transient due to the use of the long sampling period at the high level.
 
For this reason, at the low level the overall control action associated with $y_f$ is refined as
\begin{subequations}\label{Eqn: u_redefine}
	\begin{align}\label{Eqn: u_f_redefine}
	u_f(h)=\bar u_f(h)+\delta u_f(h)
	\end{align}
	where $\bar u_f(h)=u_f^{\scriptscriptstyle[N]}(\lfloor{h/N}\rfloor)$, $\delta u_f$ is computed at the low level by a properly defined optimization problem, which is deferred in \eqref{Eqn:LLoptimiz}.\\
	Since $\delta u_f(h)$  could influence the value of $y_s(h)$ in the fast time scale due to possible nonzero coupling terms from $\Sigma_f$ to $\Sigma_s$ (e.g. $A_{sf}$, $B_{sf}$), it is also  convenient to allow a further control freedom of $u_s$ leading to the correction as follows:
	\begin{align}\label{Eqn: u_s_redefine}
	u_s(h)=\bar u_s(h)+\delta u_s(h)
	\end{align}
\end{subequations}
where $\bar u_s(h)=u_s^{\scriptscriptstyle[N]}(\lfloor{h/N}\rfloor)$, $\delta u_s$ is another decision variable at the low level.\\
	{\color{black}In view of~\eqref{Eqn: u_redefine}, one can rewrite~\eqref{Eqn:CL} with $u(h)=\bar u(h)+\delta u(h)$, where $\bar u=(\bar u_s,\ \bar u_f)$, $\delta u=(\delta u_s,\ \delta u_f)$. However, for prediction purpose, we define a prediction model by neglecting the effect of $d$:
\begin{equation}\label{Eqn:CL_final}
\hat \Sigma:\ \left\{\begin{array}{l}
\hat x(h+i+1|h)=A\hat x(h+i|h)+B u(h+i|h)\\
[0.2cm]\hat y(h+i|h)=C\hat x(h+i|h)
\end{array}\right. \qquad
\end{equation}
where $\hat x(h|h)=x(h)$.}
Accordingly, at any fast time instant $h=kN+t$, letting $\overrightarrow{\delta{u}}(h:(k+1)N-1)=\big[\, \delta u(h|h)\ \cdots\ \delta u((k+1)N-1|h)\,\big]\in({\mathbb R}^{m})^{N-t}$, a shrinking horizon MPC problem  can be solved at the low level:
 
\begin{equation}\label{Eqn:LLoptimiz}
\begin{array}{ccl}
 
\min&  {J_{\rm\scriptscriptstyle L}}\\
{\scriptstyle\overrightarrow{\delta {u}}{\scriptscriptstyle(h:(k+1)N-1)}}&
\end{array}
\end{equation}
where
\begin{equation}\label{Eqn:LLoptimiz_cost}
J_{\rm\scriptscriptstyle L}=\sum_{j=0}^{N-t-1}\|\hat y(h+j|h)-\tilde y^\ast(h)\|_{Q}^2+\| \delta u(h+j|h)\|_{R}^2
\end{equation}
and where $\tilde y^\ast(h)=(\tilde y_s(h),\ \tilde y_f(kN+N))$, $h\in[kN,kN+N)$, $\tilde y_s$ and $\tilde y_f$ are defined above~\eqref{Eqn: u_s_redefine}.\\
The optimization problem~\eqref{Eqn:LLoptimiz} is performed under the following constraints:
\begin{enumerate} [(1)]
	\item the dynamics~\eqref{Eqn:CL_final};
		{\color{black}
	\item the input constraint
	\begin{equation}\label{Eqn:LLoptimiz1_con}
   \begin{array}{ll}
		u^{\scriptscriptstyle[N]}(\lfloor{h/N}\rfloor)+\delta u(h|h)\in{\mathcal U},\\
		u^{\scriptscriptstyle[N]}(\lfloor{h/N}\rfloor)+\delta u(h+j|h)\in\hat{\mathcal U}^i,\ \forall\,j=1,\dots,N-t-1,
\end{array}
	\end{equation}
	where $\hat{\mathcal U}^i$ is such that $\bar{\mathcal U}\subseteq\hat{\mathcal U}^i\subseteq{\mathcal U}$, $i=h-kN$;
	\item the state terminal constraint
	\begin{equation}\label{LLoptimiz_term_con}
	\hat x(kN+N|h)= \hat x^{\scriptscriptstyle[N]}(k+1|k).
	\end{equation}}
\end{enumerate}
\begin{remark}
	The rationale of choosing signal $\tilde y^\ast(h)$ as the reference for the low level lies in the fact that  $\hat y_f(h)$ is expected to react promptly to respond to $\tilde y_f(kN+N)$, while $\hat y_s(h)$ can be controlled to follow the smooth trajectory $\tilde y_s(h)$ generated from the high level.
\end{remark}
{\color{black}
	\begin{remark}
		It is highlighted that the structure of the proposed approach is different from that of the cascade ones, see for instance~\cite{picasso2010mpc}, for the reason that in the cascade algorithm, the computed input from the high level is considered as the output reference to be tracked at the low level, while the proposed algorithms utilize the control action computed from the high level, i.e., $\bar u(h)$, to generate the possible reference profile with model~\eqref{Eqn:CL} in an open-loop fashion.
	\end{remark}
}
\subsection{Summary of the D-MPC algorithm}
In summary, the main steps for the on-line implementation of the D-MPC are given in Algorithm \ref{Atm:1}.
\begin{algorithm}
	initialization\;\\
	\While{ for any integer $k\geq0$}{
		\textbf{h1)} compute the control $u^{{\scriptscriptstyle[N]}}(k|k)$ by solving the optimization problem \eqref{Eqn:HLoptimiz} with \eqref{Eqn:HL_cost} and 	{\color{black}update $\hat x^{\scriptscriptstyle[N]}(k+1|k)$} \;
		
		\textbf{h2)} generate the open-loop output $\tilde{y}^{\ast}(h)$ from \eqref{Eqn:CL} with $u(h)=u^{\scriptscriptstyle[N]}(\lfloor h/N\rfloor)$, for all $h\in[kN,kN+N)$\\
		\For{$h\leftarrow kN$ \KwTo $kN+N-1$}{

			\textbf{l1)} compute $\delta u(h|h)$ with optimization problem \eqref{Eqn:LLoptimiz} with \eqref{Eqn:LLoptimiz_cost}
			and apply  $u(h)=u^{\scriptscriptstyle[N]}(\lfloor h/N\rfloor)+\delta u(h|h)$ to~\eqref{Eqn:CL}
			
		{\color{black}
			\textbf{l2)} update $x(h+1)$, $y(h+1)$, set $\hat x(h+1)=x(h+1)$, $\hat y(h+1)=y(h+1)$
		}}
		\textbf{h3)} $k\leftarrow k+1$
	}
	\caption{On-line implementation of D-MPC}\label{Atm:1}
\end{algorithm}
 
\begin{theorem}\label{theorem}
	Under Assumption \ref{assump:A}, 
	if the initial condition is such that $\hat x^{\scriptscriptstyle[N]}(0)=\hat x(0)=x(0)$ and~\eqref{Eqn:HLoptimiz} is feasible at $k=0$, then the following results can be stated for the proposed D-MPC control algorithm.
	\begin{enumerate} [(1)]
			{\color{black}
		\item  Provided that
		\begin{equation}\label{Eqn:state-terminal-size-dmpc}
		  ( A^{\scriptscriptstyle[N]})^{ N_{\rm\scriptscriptstyle H}-1}\mathcal{D}\subseteq {\mathcal{X}}_{\rm\scriptscriptstyle F}\ominus{\mathcal{X}}_{\rm\scriptscriptstyle F}^s
		\end{equation}
		and
		at any time $h\in[kN,kN+N-2]$ that,
		\begin{equation}\label{Eqn:fea-con-dmpc}
	-A^{N-j-1}\mathcal{D}\oplus  L^j \hat{\mathcal{U}}^j\subseteq  \Delta L^j {\mathcal{U}}\oplus L^{j+1} \hat{\mathcal{U}}^{j+1}
		\end{equation}
		where $L^j=\sum_{i=j+1}^{N-1}A^{N-i-1}B$, $\Delta L^j=L^{j}-L^{j+1}$, $j=h-kN$, 
	}
		then the feasibility can be guaranteed:
		\begin{itemize}
			\item for the high-level problem \eqref{Eqn:HLoptimiz} at all slow time instant $k>0$;
			\item for the low-level problem \eqref{Eqn:LLoptimiz}  at all fast time instant $h\geq 0$.
		\end{itemize}
   \item  	{\color{black}If the disturbance $d=0$, then the asymptotic convergence of the closed-loop system can be ensured:}
   \begin{itemize}
       \item The  slow-time scale system $\Sigma^{\scriptscriptstyle[N]}$ enjoys the convergence property, i.e.,
		$\lim_{k\to+\infty} (u^{\scriptscriptstyle[N]}(k),\, x^{\scriptscriptstyle[N]}(k),\, y^{\scriptscriptstyle[N]}(k))=(u_r,\, x_r,\, y_r)$.
		\item Moreover, for the low-level problem \eqref{Eqn:LLoptimiz}, it holds that $\lim_{h\to+\infty} \delta u(h)=0$. Finally,
		$\lim_{h\to+\infty}(u(h),\, x(h),\, y(h))=(u_r,\, x_r,\, y_r)$.
   \end{itemize}
		
	\end{enumerate}
\end{theorem}
Note that, the terminal constraint~\eqref{LLoptimiz_term_con} plays a crucial role for guaranteeing the closed-loop properties of the D-MPC.
However, as the proposed D-MPC control structure is an upper-bottom one, the computed value of $\hat x_f^{\scriptscriptstyle[N]}(k)$ at the high level influences the control performance at the low level due to~\eqref{LLoptimiz_term_con}.  For this reason, the state $\hat x_f(h)$ associated with $\hat y_f(h)$ in the basic time scale might not converge to its nominal value faster than $\hat x_s(h)$ especially for systems that exhibit nonseparable open-loop dynamics.
We solve this problem in the following section.
 
\section{\emph{Incremental} D-MPC algorithm}\label{sec:iD-MPC}
{\color{black}
In this section, we design an \emph{Incremental} D-MPC algorithm to improve the control performance associated with the ``fast" output $y_f$ and to compensate for possible time-varying piece-wise constant or smooth uncertainties.}
%
\subsection{Design of the Incremental D-MPC}
In the following we first focus on the re-design of the MPC regulator at the high level.
To this end, we first partition the above sampled system as the one with the structure similar to~\eqref{Eqn:sigma_s_f}.
{\color{black}To proceed, we  rewrite
the matrices $A^{\scriptscriptstyle[N]}$, $B^{\scriptscriptstyle[N]}$, and vector $\tilde d^{\scriptscriptstyle[N]}$ into the following forms
$$
A^{\scriptscriptstyle[N]}=\begin{bmatrix}A_{ss}^{\scriptscriptstyle[N]}&A_{sf}^{\scriptscriptstyle[N]}\\A_{fs}^{\scriptscriptstyle[N]}&A_{ff}^{\scriptscriptstyle[N]}\end{bmatrix},\ \ B^{\scriptscriptstyle[N]}=\begin{bmatrix}B_{ss}^{\scriptscriptstyle[N]}&B_{sf}^{\scriptscriptstyle[N]}\\B_{fs}^{\scriptscriptstyle[N]}&B_{ff}^{\scriptscriptstyle[N]}\end{bmatrix},\ \
\tilde d^{\scriptscriptstyle[N]}=\begin{bmatrix}
\tilde d_s^{\scriptscriptstyle[N]}\\\tilde d_f^{\scriptscriptstyle[N]}
\end{bmatrix}$$
where $A_{ss}^{\scriptscriptstyle[N]}\in\mathbb{R}^{n_s\times n_s}$, $B_{ss}^{\scriptscriptstyle[N]}\in\mathbb{R}^{n_s\times m_s}$, $\tilde d_s^{\scriptscriptstyle[N]}\in \mathbb{R}^{n_s}$.

In view of this, the sampled perturbed system~\eqref{Eqn:CL_k} can be partitioned as two interacting ones as follows:
\begin{subequations}\label{Eqn:sigma_s_f-k-im}
	\begin{align}
	&\Sigma_s^{\scriptscriptstyle[N]}:\ \left\{\begin{array}{ll}
	x_s^{\scriptscriptstyle[N]}(k+1)=&A_{ss}^{\scriptscriptstyle[N]}x_s^{\scriptscriptstyle[N]}(k)+A_{sf}^{\scriptscriptstyle[N]}x_f^{\scriptscriptstyle[N]}(k)+B_{ss}^{\scriptscriptstyle[N]}u_s^{\scriptscriptstyle[N]}(k)\\&+B_{sf}^{\scriptscriptstyle[N]}u_f^{\scriptscriptstyle[N]}(k)+\tilde d^{\scriptscriptstyle[N]}_s(k)\\
	[0.2cm]y_s^{\scriptscriptstyle[N]}(k)=&C_{ss}x_s^{\scriptscriptstyle[N]}(k),
	\end{array} \right.\qquad\label{Eqn:sigma_s-k}\\
	&\Sigma_f^{\scriptscriptstyle[N]}:\ \left\{\begin{array}{ll}
	x_f^{\scriptscriptstyle[N]}(k+1)=&A_{fs}^{\scriptscriptstyle[N]}x_s^{\scriptscriptstyle[N]}(k)+A_{ff}^{\scriptscriptstyle[N]}x_f^{\scriptscriptstyle[N]}(k)+B_{fs}^{\scriptscriptstyle[N]}u_s^{\scriptscriptstyle[N]}(k)\\
	&+B_{ff}^{\scriptscriptstyle[N]}u_f^{\scriptscriptstyle[N]}(k)+\tilde d^{\scriptscriptstyle[N]}_f(k)\\
	[0.2cm]y_f^{\scriptscriptstyle[N]}(k)=&C_{ff}x_f^{\scriptscriptstyle[N]}(k),
	\end{array} \right.\qquad\label{Eqn:sigma_f-k}
	\end{align}
\end{subequations}
}
The following assumption about $\Sigma_f^{\scriptscriptstyle[N]}$ is assumed to hold:
\begin{assumption}\label{assum:cb-fullrank}
	Matrix $C_{ff}B_{ff}^{\scriptscriptstyle[N]}$ is full rank.
\end{assumption}
{\color{black}With~\eqref{Eqn:sigma_s_f-k-im}, with the goal of guaranteeing satisfactory control performance related to $y_f$ in the basic time scale, it is convenient to enforce all the future predictions $y_f^{\scriptscriptstyle[N]}(k),\,\forall k>0$ associated with  $\Sigma_f^{\scriptscriptstyle[N]}$ being equal to the reference value $y_{f,r}$. In this way, the real output $y_f$ resulting from controller~\eqref{Eqn:LLoptimiz}  will reach the reference $y_{f,r}$ in only one slow-time step.}  However, this restriction might cause infeasibility issue in case $y_{f,r}$ is far from its initial value $y_f(0)$, and/or constraints on the control increments are enforced. 
For this reason, instead of imposing $y_f^{\scriptscriptstyle[N]}(k)=y_{f,r}\ \forall k>0$, one can enforce the following relation
\begin{equation}\label{Eqn:yf-gov}
y_f^{\scriptscriptstyle[N]}(k)=\tilde y_{f,r}(k)\ \forall k>0,
\end{equation}
where $\tilde y_{f,r}(k)=y_f(0)+\alpha(k)(y_{f,r}-y_f(0))$ and where $\alpha(k)$ is defined as an optimization variable that its value is restricted by $0\leq\alpha(k)\leq1$  and reaches $1$ in finite time steps, i.e.,
\begin{equation}\label{Eqn:alpha}
\left\{\begin{array}{ll}
\alpha(0)=0,& \\
0\leq\alpha(k)\leq 1,& k\in[1,\ N_{\alpha})\\
\alpha(k)=1,& k\geq N_{\alpha}
\end{array}\right.
\end{equation}
where $N_{\alpha}$ is a positive integer. {\color{black}As $\tilde d^{\scriptscriptstyle[N]}_f(k)$ is unknown at time $k$,~\eqref{Eqn:yf-gov} can be slightly relaxed, i.e., we enforce  at time $k$  $\hat y_f^{\scriptscriptstyle[N]}(k+j)=\tilde y_{f,r}(k+j)$. In view of~\eqref{Eqn:sigma_s_f-k-im}, it is required that
\begin{equation}\label{Eqn:uf-yf-con-k-im}
\begin{array}{ll}
u_f^{\scriptscriptstyle[N]}(k+j)=&
(C_{ff}B_{ff}^{\scriptscriptstyle[N]})^{-1}\big( \tilde y_{f,r}(k+j)-\\ &C_{ff}(\begin{bmatrix}
A_{fs}^{\scriptscriptstyle[N]}&A_{ff}^{\scriptscriptstyle[N]}
\end{bmatrix}\hat x^{\scriptscriptstyle[N]}(k+j)+B_{fs}^{\scriptscriptstyle[N]}u_s^{\scriptscriptstyle[N]}(k+j))\big)
\end{array}
\end{equation}
where $\hat x$ is the predicted value of $x$.}
Under constraint~\eqref{Eqn:uf-yf-con-k-im}, the time steps required for  $\hat y_f^{\scriptscriptstyle[N]}=y_{f,r}$ can be defined via properly tuning parameter $N_{\alpha}$.
 
 By substituting $u_f^{\scriptscriptstyle[N]}$ with~\eqref{Eqn:uf-yf-con-k-im}, one can write the one-step ahead state prediction at time $k$, i.e.,
\begin{equation}\label{Eqn:sigma_s-us-k-im}
 \left\{\begin{array}{l}
\hat x^{\scriptscriptstyle[N]}(k+1)=\tilde A^{\scriptscriptstyle[N]}\hat x^{\scriptscriptstyle[N]}(k)+\tilde B_{s}^{\scriptscriptstyle[N]}
u_s^{\scriptscriptstyle[N]}(k)+\tilde B_{f}^{\scriptscriptstyle[N]}\tilde y_{f,r}(k)\\
[0.2cm]\hat y_s^{\scriptscriptstyle[N]}(k)=\tilde C_{s}\hat x^{\scriptscriptstyle[N]}(k),
\end{array} \right.\qquad
\end{equation}
where 
$\hat x^{\scriptscriptstyle[N]}(k)= x^{\scriptscriptstyle[N]}(k)$, 
$\tilde A^{\scriptscriptstyle[N]}=\begin{bmatrix}\tilde A_{ss}^{\scriptscriptstyle[N]}&\tilde A_{sf}^{\scriptscriptstyle[N]}\\ \tilde A_{fs}^{\scriptscriptstyle[N]}&\tilde A_{ff}^{\scriptscriptstyle[N]}\end{bmatrix}$, $\tilde B_s^{\scriptscriptstyle[N]}=\begin{bmatrix}\tilde B_{ss}^{\scriptscriptstyle[N]}\\\tilde B_{fs}^{\scriptscriptstyle[N]}\end{bmatrix}$, $\tilde B_f^{\scriptscriptstyle[N]}=\begin{bmatrix}\tilde B_{sf}^{\scriptscriptstyle[N]}\\\tilde B_{ff}^{\scriptscriptstyle[N]}\end{bmatrix}$, $\tilde C_s=\begin{bmatrix}C_{ss}\\0\end{bmatrix}^{\top}$, and where
\begin{align*}
&\tilde A_{ss}^{\scriptscriptstyle[N]}=A_{ss}^{\scriptscriptstyle[N]}-B_{sf}^{\scriptscriptstyle[N]}(C_{ff}B_{ff}^{\scriptscriptstyle[N]})^{-1}C_{ff}A_{fs}^{\scriptscriptstyle[N]}\\
&\tilde A_{sf}^{\scriptscriptstyle[N]}=A_{sf}^{\scriptscriptstyle[N]}-B_{sf}^{\scriptscriptstyle[N]}(C_{ff}B_{ff}^{\scriptscriptstyle[N]})^{-1}C_{ff}A_{ff}^{\scriptscriptstyle[N]}\\
&\tilde A_{fs}^{\scriptscriptstyle[N]}=A_{fs}^{\scriptscriptstyle[N]}-B_{ff}^{\scriptscriptstyle[N]}(C_{ff}B_{ff}^{\scriptscriptstyle[N]})^{-1}C_{ff}A_{fs}^{\scriptscriptstyle[N]}\\
&\tilde A_{ff}^{\scriptscriptstyle[N]}=A_{ff}^{\scriptscriptstyle[N]}-B_{ff}^{\scriptscriptstyle[N]}(C_{ff}B_{ff}^{\scriptscriptstyle[N]})^{-1}C_{ff}A_{ff}^{\scriptscriptstyle[N]}\\
&\tilde B_{ss}^{\scriptscriptstyle[N]}=B_{ss}^{\scriptscriptstyle[N]}-B_{sf}^{\scriptscriptstyle[N]}(C_{ff}B_{ff}^{\scriptscriptstyle[N]})^{-1}C_{ff}B_{fs}^{\scriptscriptstyle[N]}\\
&\tilde B_{fs}^{\scriptscriptstyle[N]}=B_{fs}^{\scriptscriptstyle[N]}-B_{ff}^{\scriptscriptstyle[N]}(C_{ff}B_{ff}^{\scriptscriptstyle[N]})^{-1}C_{ff}B_{fs}^{\scriptscriptstyle[N]}\\
&\tilde B_{sf}^{\scriptscriptstyle[N]}=B_{sf}^{\scriptscriptstyle[N]}(C_{ff}B_{ff}^{\scriptscriptstyle[N]})^{-1}\\
&\tilde B_{ff}^{\scriptscriptstyle[N]}=B_{ff}^{\scriptscriptstyle[N]}(C_{ff}B_{ff}^{\scriptscriptstyle[N]})^{-1}.
\end{align*}
\begin{assumption}\label{assum:tilde A-stable}
	The integer $N$ is such that $\tilde A^{\scriptscriptstyle[N]}$ is stabilizable.
\end{assumption}
{\color{black} 
	To account for the model uncertainties,
	model~\eqref{Eqn:sigma_s-us-k-im} is reformulated in the \emph{incremental form} and used for defining an MPC including an integral action. 
	{\color{black}In doing so, the closed-loop system is capable to compensate for the influences caused by smoothly varying disturbances, see~\cite{betti}.} To this end,
letting
 $\bar {\hat x}^{\scriptscriptstyle[N]}(k)=(\hat y_s^{\scriptscriptstyle[N]}(k),\ \Delta {\hat x}^{\scriptscriptstyle[N]}(k))$, from~\eqref{Eqn:sigma_s-us-k-im} we compute
\begin{equation}\label{Eqn:sigma_s-dus-k}
 \left\{\begin{array}{rl}
\bar {\hat x}^{\scriptscriptstyle[N]}(k+1)=&\bar A^{\scriptscriptstyle[N]}\bar {\hat x}^{\scriptscriptstyle[N]}(k)+\bar B_{s}^{\scriptscriptstyle[N]}
\Delta u_s^{\scriptscriptstyle[N]}(k)+\\
&\bar B_{f}^{\scriptscriptstyle[N]}\Delta \alpha(k)(y_{f,r}-y_f(0))\\
\alpha(k)=&\alpha(k-1)+\Delta\alpha(k)\\
[0.2cm]\hat y_s^{\scriptscriptstyle[N]}(k)=&\bar C\bar {\hat x}^{\scriptscriptstyle[N]}(k),
\end{array} \right.\qquad
\end{equation}
where $\bar A^{\scriptscriptstyle[N]}=\begin{bmatrix}I& \tilde C_s\tilde A^{\scriptscriptstyle[N]}\\0&\tilde A^{\scriptscriptstyle[N]}\end{bmatrix}$, $\bar B_s^{\scriptscriptstyle[N]}=\begin{bmatrix}\tilde C_s\tilde B_s^{\scriptscriptstyle[N]}\\ \tilde B_s^{\scriptscriptstyle[N]}\end{bmatrix}$, $\bar B_f^{\scriptscriptstyle[N]}=\begin{bmatrix}\tilde C_s\tilde B_f^{\scriptscriptstyle[N]}\\ \tilde B_f^{\scriptscriptstyle[N]}\end{bmatrix}$, $\bar C=\begin{bmatrix}I&0\end{bmatrix}$. 
\begin{proposition}\label{prop1}
	The pair $(\bar A^{\scriptscriptstyle[N]},\ \bar B_s^{\scriptscriptstyle[N]})$ is stabilizable if and only if
	$$\bullet\ \ \text{rank}\big(\begin{bmatrix}
	\tilde C_s\tilde A^{\scriptscriptstyle[N]}&\tilde C_s\tilde B_s^{\scriptscriptstyle[N]}\\\tilde A^{\scriptscriptstyle[N]}-I&\tilde B_s^{\scriptscriptstyle[N]}
	\end{bmatrix}^{\top} \big)= n+p_s,$$
	$$\bullet\ \ \text{rank}\big(\begin{bmatrix}
	2I&0\\
	\tilde C_s\tilde A^{\scriptscriptstyle[N]}&\tilde A^{\scriptscriptstyle[N]}+I\\\tilde C_s\tilde B_s^{\scriptscriptstyle[N]}&\tilde B_s^{\scriptscriptstyle[N]}
	\end{bmatrix}^{\top}\big)= n+p_s.$$
\end{proposition}
Under proposition~\ref{prop1}, it is possible to find a gain matrix $\bar K_{s,\rm\scriptscriptstyle H}$ such that $\bar F_{s,\rm\scriptscriptstyle H}=\bar A^{\scriptscriptstyle[N]}+\bar B^{\scriptscriptstyle[N]}\bar K_{s,\rm\scriptscriptstyle H}$ is Schur stable.
 
%
{\color{black}
Note that, it is not straightforward to write  the constraints on 
$\bar  u_s^{\scriptscriptstyle[N]}$ and $\bar  u_f^{\scriptscriptstyle[N]}$ using model~\eqref{Eqn:sigma_s-dus-k}.  We are going to show that, in line with~\cite{betti}, it is possible to represent the control variables by the state in the \emph{inremental form}, i.e.,
\begin{equation}\label{Eqn:con-velocity}
	\begin{array}{ll}
 \bar u_s^{\scriptscriptstyle[N]}(k)=& \Gamma_{us}(\bar {\hat x}^{\scriptscriptstyle[N]}(k+1)-\bar B_f^{\scriptscriptstyle[N]}\tilde y_{f,r})\\
\bar	u_f^{\scriptscriptstyle[N]}(k)=&(C_{ff}B_{ff}^{\scriptscriptstyle[N]})^{-1}(\tilde y_{f,r}(k)-\\
	&\Gamma_{uf}(\bar {\hat x}^{\scriptscriptstyle[N]}(k+1)-\bar B_f^{\scriptscriptstyle[N]}\tilde y_{f,r}))
	\end{array}
\end{equation}
where $\Gamma_{uf}=
C_{ff}\begin{bmatrix}A_{fs}^{\scriptscriptstyle[N]}&A^{\scriptscriptstyle[N]}_{ff}&B_{fs}^{\scriptscriptstyle[N]}
\end{bmatrix}\Gamma$, 
$\Gamma_{us}=\begin{bmatrix}
0_n&I_{m_s}
\end{bmatrix}\Gamma$,  and where $\Gamma=\begin{bmatrix}
\tilde C_s\tilde A^{\scriptscriptstyle[N]}&\tilde C_s\tilde B_s^{\scriptscriptstyle[N]}\\\tilde A^{\scriptscriptstyle[N]}-I_{n}&\tilde B_s^{\scriptscriptstyle[N]}
\end{bmatrix}^{-1}$.\\
In view of~\eqref{eq:u_s_constraint},~\eqref{eq:u_f_constraint},~\eqref{Eqn:con-velocity}, to enforce constraints on $(\bar u_s^{\scriptscriptstyle[N]}(k), \bar u_f^{\scriptscriptstyle[N]}(k))
\in \mathcal{U}_s\times\mathcal{U}_f$, one can use 
\begin{equation}\label{Eqn:con-velocity-all}
\begin{array}{c}
A_x\bar x^{\scriptscriptstyle[N]}(k+1)+b_y\in \bar{\mathcal{U}} 
\end{array}
\end{equation}
where 
$A_x=\begin{bmatrix}
\Gamma_{us}^{\top}&-\Gamma_{uf}^{\top}\end{bmatrix}^{\top}$, $b_y=\begin{bmatrix}-(\Gamma_{us}\bar B_f^{\scriptscriptstyle[N]})^{\top} &(C_{ff}B_{ff}^{\scriptscriptstyle[N]})^{-\top}+((C_{ff}B_{ff}^{\scriptscriptstyle[N]})^{-1}\Gamma_{uf}\bar B_f^{\scriptscriptstyle[N]})^{\top}\end{bmatrix}^{\top}\tilde y_{f,r}$. 
}
 
%
 
Based on ~\eqref{Eqn:sigma_s-dus-k} and~\eqref{Eqn:con-velocity-all}, now we state the \emph{Incremental} MPC problem at the high level. 
At each slow time-step $k$ we solve an optimization problem according to receding horizon principle as follows:
\begin{equation}
\begin{array}{cc}
\min & {\bar J_{{\rm\scriptscriptstyle H}}}\\
 
{\scriptstyle \overrightarrow{\Delta u_s^{\scriptscriptstyle[N]}}{(k:k+\bar N_{\rm\scriptscriptstyle H}-1)}}\\
\end{array}\label{Eqn:HLoptimiz-im}
\end{equation}
where
\begin{equation}
\begin{array}{cll}
\bar J_{{\rm\scriptscriptstyle H}}=&\sum_{i=0}^{\bar N_{\rm\scriptscriptstyle H}-1}\|\bar {\hat x}^{\scriptscriptstyle[N]}(k+i)-\bar C^{\top} y_{s,r}\|_{\bar Q_{s,\rm\scriptscriptstyle H}}^{2}+\|\Delta u_s^{\scriptscriptstyle[N]}(k+i)\|_{\bar R_{s,\rm\scriptscriptstyle H}}^{2}+\\
&\gamma(\alpha(k+i)-1)^2+ \|\bar {\hat x}^{\scriptscriptstyle[N]}(k+\bar N_{\rm\scriptscriptstyle H})-\bar C^{\top}y_{s,r}\|_{\bar P_{\rm\scriptscriptstyle H}}^{2}.
\end{array}\label{Eqn:HL_cost-im}
\end{equation}	
$\gamma$ is a positive scalar, $\bar N_{\rm\scriptscriptstyle H}>N_{\alpha}$ is the adopted prediction horizon.
The positive definite and symmetric weighting matrices ${\bar Q_{s,\rm\scriptscriptstyle H}}\in \mathbb{R}^{(n+p_s)\times (n+p_s)}$, ${\bar R_{s,\rm\scriptscriptstyle H}}\in\mathbb{R}^{m_s\times m_s}$ are free design parameters, while  ${\bar P_{\rm\scriptscriptstyle H}}$ is computed as the solution to the Lyapunov equation
\begin{equation}\label{Eqn:lyap-im}
\bar F_{s,\rm\scriptscriptstyle H}^{\top} \bar P_{\rm\scriptscriptstyle H} \bar F_{s,\rm\scriptscriptstyle H}-\bar P_{\rm\scriptscriptstyle H}=-(Q_{s,\rm\scriptscriptstyle H}+\bar K_{s,\rm\scriptscriptstyle H}^{\top} \bar R_{s,\rm\scriptscriptstyle H} \bar K_{s,\rm\scriptscriptstyle H})
\end{equation}
The optimization problem~\eqref{Eqn:HLoptimiz-im} is performed under the following constraints:
	{\color{black}
\begin{enumerate}[(1)]
	\item  dynamics~\eqref{Eqn:sigma_s-dus-k}, constraint~\eqref{Eqn:alpha}, and~\eqref{Eqn:con-velocity-all};
	\item the state terminal constraint
	\begin{equation*}
	\begin{array}{ccc}
	\bar x^{\scriptscriptstyle[N]}(k+\bar N_{\rm\scriptscriptstyle H})\in \bar{\mathcal{X}}_{\rm\scriptscriptstyle F}^s,\\
	\end{array}
	\end{equation*}
\end{enumerate}
where $\bar{\mathcal{X}}_{\rm\scriptscriptstyle F}^s\subseteq\bar{\mathcal{X}}_{\rm\scriptscriptstyle F}$,}  the set $\bar{\mathcal{X}}_{\rm\scriptscriptstyle F}$ is a positively invariant set for the nominal system of~\eqref{Eqn:sigma_s-dus-k}, i.e.,
\begin{equation}\label{Eqn:sigma_s-dus-k-nominal}
\begin{array}{l}
z(k+1)=\bar A^{\scriptscriptstyle[N]}z(k)+\bar B_{s}^{\scriptscriptstyle[N]}
\hat u(k),
\end{array}\qquad
\end{equation}
that is controlled with the stabilizing control law $\hat u(k)=\bar K_{s,\rm\scriptscriptstyle H} (z(k)-\bar C^{\top}y_{s,r})$ such that $\bar F_{s,\rm\scriptscriptstyle H}\bar{\mathcal{X}}_{\rm\scriptscriptstyle F}\subseteq \bar{\mathcal{X}}_{\rm\scriptscriptstyle F}$ under constraint~\eqref{Eqn:con-velocity-all}.  	{\color{black}The set $\bar{\mathcal{X}}_{\rm\scriptscriptstyle F}^s$ is selected such that: for any $z(k)\in \bar{\mathcal{X}}_{\rm\scriptscriptstyle F}$, the successive state under the above stabilizing control law satisfies  $z(k+1)\in \bar{\mathcal{X}}_{\rm\scriptscriptstyle F}^s$.
}

Let $\overrightarrow{\Delta u_s^{\scriptscriptstyle[N]}}{(k:k+\bar N_{\rm\scriptscriptstyle H}-1|k)}$ be the optimal solution to optimization \eqref{Eqn:HLoptimiz-im}. 
The real input $u_s^{\scriptscriptstyle[N]}(k)$ at time instant $k$ is given by $u_s^{\scriptscriptstyle[N]}(k)=u_s^{\scriptscriptstyle[N]}(k-1)+\Delta u_s^{\scriptscriptstyle[N]}(k|k)$, also  from~\eqref{Eqn:uf-yf-con-k-im} we can compute the value of $u_f^{\scriptscriptstyle[N]}(k)$. At this time instant, the state $\hat x^{\scriptscriptstyle[N]}(k+1|k)$ is available by applying $u^{\scriptscriptstyle[N]}(k)=(u_s^{\scriptscriptstyle[N]}(k),u_f^{\scriptscriptstyle[N]}(k))$ to \eqref{Eqn:CL_k}.
 
{\color{black}
In principle, the fast MPC problem described in the previous section, i.e., \eqref{Eqn:LLoptimiz} with cost~\eqref{Eqn:LLoptimiz1_con}, can be used for computing the corrections of the control input in the fast time scale. We propose an improved algorithm in the following. Slightly different to~\eqref{Eqn:CL_final} in \eqref{Eqn:LLoptimiz}, we use 
\begin{equation}\label{Eqn:CL_final-disturbance}
 \left\{\begin{array}{ll}
\Delta\hat x(h+i+1|h)=A\Delta\hat x(h+i|h)+B\Delta u(h+i|h)\\
[0.2cm]\hat y(h+i|h)=\hat y(h+i-1|h)+C\Delta\hat x(h+i|h)
\end{array}\right. \qquad
\end{equation}
 $h\in [kN,\ kN+N)$, to compensate for uncertainties.
%
 
Accordingly, at any fast time instant $h=kN+t$, letting $\overrightarrow{\Delta{u}}(h:(k+1)N-1)=\big[\, \Delta u(h|h)\ \cdots\ \Delta u((k+1)N-1|h)\,\big]\in({\mathbb R}^{m})^{N-t}$, $\bar {\hat x}=(\hat y,\Delta \hat x)$, and $\bar y=(\tilde y^\ast,0)$, the shrinking horizon MPC problem  can be solved at the low level:
 
\begin{equation}\label{Eqn:LLoptimiz_rev}
\begin{array}{ccl}
 
\min&  {\bar J_{\rm\scriptscriptstyle L}}\\
{\scriptstyle\overrightarrow{\Delta {u}}{\scriptscriptstyle(h:(k+1)N-1)}}&
\end{array}
\end{equation}
\begin{equation}\label{Eqn:LLoptimiz_cost_rev}
\bar J_{\rm\scriptscriptstyle L}=\sum_{j=0}^{N-t-1}\|\bar {\hat x}(h+j|h)-\bar y(h)\|_{\bar Q}^2+\| \Delta u(h+j|h)\|_{R}^2
\end{equation}
where $\bar Q\in\mathbb{R}^{(n+p)\times (n+p)}$ is a positive-definite matrix.}
The optimization problem~\eqref{Eqn:LLoptimiz_rev} is performed under the following constraints:
\begin{enumerate} [(1)]
	\item the dynamics~\eqref{Eqn:CL_final-disturbance};
		{\color{black}
	\item the input constraint
	\begin{equation}\label{Eqn:LLoptimiz1_con_rev}
\begin{array}{ll}
	\hspace{-3mm}	u(kN-1)+ \Delta u(kN|kN)\in {\mathcal U},\\
	\hspace{-3mm}		u(kN-1)+\sum_{i=0}^j\Delta u(h+j|h)\in\hat {\mathcal U}^i,\ \forall\,j=1,\dots,N-t-1,
		\end{array}
	\end{equation}
	where $i=h-kN$;
	\item the state terminal constraint
	\begin{equation}\label{LLoptimiz_term_con_rev}
	\hat x(kN+N|h)= \hat x^{\scriptscriptstyle[N]}(k+1|k)
	\end{equation}}
\end{enumerate}
	{\color{black}It is noted that constraints~\eqref{Eqn:LLoptimiz1_con_rev} and~\eqref{LLoptimiz_term_con_rev} can be rewritten following the line of~\eqref{Eqn:con-velocity-all}, but the design steps are neglected for the sake of simplicity.
}
{\color{black}
}
 
\subsection{Summary of the Incremental D-MPC algorithm}
To better clarify the requirements for implementing \emph{Incremental} D-MPC and its difference with the D-MPC, the main steps for the on-line implementation are given in Algorithm~\ref{Atm:2}.
\begin{algorithm}
	initialization with $N_{\alpha}=1$\;\\
	\While{ for any integer $k\geq0$}{
		\textbf{h1)} compute the control $\Delta u^{{\scriptscriptstyle[N]}}_s(k|k)$ by solving the optimization problem \eqref{Eqn:HLoptimiz-im} with \eqref{Eqn:HL_cost-im} and update $\bar x^{\scriptscriptstyle[N]}(k+1|k)$ \;\\
		\eIf{\eqref{Eqn:HLoptimiz-im} with \eqref{Eqn:HL_cost-im} is infeasible }{
			$N_{\alpha}\ \leftarrow \ N_{\alpha}+1$ and go back to step \textbf{h1)}, (see~\eqref{Eqn:alpha}) \;
		}{
			continue\;
		}
		\textbf{h2)} calculate $u_f^{{\scriptscriptstyle[N]}}(k)$ from~\eqref{Eqn:uf-yf-con-k-im} with $u_s^{{\scriptscriptstyle[N]}}(k)=u_s^{{\scriptscriptstyle[N]}}(k-1)+\Delta u_s^{{\scriptscriptstyle[N]}}(k|k)$, apply the control $u^{{\scriptscriptstyle[N]}}(k)=(u_s^{{\scriptscriptstyle[N]}}(k),u_f^{{\scriptscriptstyle[N]}}(k))$ to \eqref{Eqn:CL_C} and update $x^{\scriptscriptstyle[N]}(k+1|k)$\;
		
		\textbf{h3)} generate the open-loop output $\tilde{y}^{\ast}(h)$ from \eqref{Eqn:CL} with $u(h)=u^{\scriptscriptstyle[N]}(\lfloor h/N\rfloor)$, for all $h\in[kN,kN+N)$\\
		\For{$h\leftarrow kN$ \KwTo $kN+N-1$}{
			
				{\color{black}
			\textbf{l1)} compute $\Delta u(h|h)$ using optimization problem \eqref{Eqn:LLoptimiz_rev} with \eqref{Eqn:LLoptimiz_cost_rev}
			and apply  $u(h)=u(h-1)+\Delta u(h|h)$ to~\eqref{Eqn:CL}
			
			\textbf{l2)} update $x(h+1)$ and $y(h+1)$, set $\hat x(h+1)=x(h+1)$, $\hat y(h+1)=y(h+1)$
		}}
		\textbf{h4)} $k\leftarrow k+1$
	}
	\caption{On-line implementation of \emph{Incremental} D-MPC}\label{Atm:2}
\end{algorithm}
 
{\color{black}
\begin{theorem}\label{theorem-im}
	Under Assumptions~\ref{assump:A}--\ref{assum:tilde A-stable}, if the initial condition is such that  $\hat x^{\scriptscriptstyle[N]}(0)=\hat x(0)=x(0)$ and $N_{\alpha}$ is reachable by Algorithm~\ref{Atm:2} such that \eqref{Eqn:HLoptimiz-im} is feasible at $k=0$, then the following results can be stated for the \emph{Incremental} D-MPC algorithm under suitable conditions. 
	\begin{enumerate}[(1)]
		\item Provided that
		\begin{equation}\label{Eqn:state-terminal-size}
		(\bar A^{\scriptscriptstyle[N]})^{\bar N_{\rm\scriptscriptstyle H}-1}E\mathcal{D}\subseteq \bar{\mathcal{X}}_{\rm\scriptscriptstyle F}\ominus\bar{\mathcal{X}}_{\rm\scriptscriptstyle F}^s
		\end{equation}
where $ E=\begin{bmatrix}\tilde C_s^{\top}&I^{\top}\end{bmatrix}^{\top}$, 
		and~\eqref{Eqn:fea-con-dmpc} is satisfied,
 then the feasibility can be guaranteed:
		
		\begin{itemize}
			\item for the high-level problem \eqref{Eqn:HLoptimiz-im} at all slow time instant $k>0$;
			\item for the low-level problem \eqref{Eqn:LLoptimiz_rev}  at all fast time instant $h\geq 0$.
		\end{itemize}
		\item  If the disturbance $d$ is constant, then the asymptotic convergence of the closed-loop system can be ensured.
		\begin{itemize}
		\item The slow-time scale system \eqref{Eqn:sigma_s-dus-k} enjoys the convergence property, i.e.,
		$\lim_{k\to+\infty} (\bar x^{\scriptscriptstyle[N]}(k),\, \Delta u_s^{\scriptscriptstyle[N]}(k))=(\bar C^{\top} y_{s,r},\, 0)$. Consequently, $\lim_{k\to+\infty} (x^{\scriptscriptstyle[N]}(k),\, u^{\scriptscriptstyle[N]}(k))=(x_r,\, u_r)$.
		\item For the low-level problem \eqref{Eqn:LLoptimiz_rev}, it holds that $\lim_{h\to+\infty} \Delta u(h)=0$. Finally,
		$\lim_{h\to+\infty} (u(h),\, x(h),\, y(h))=(u_r,\, x_r,\, y_r)$.
	\end{itemize}
	\end{enumerate}
	
\end{theorem}
}
\section{Simulation example}\label{sec:example}
 In this section, simulation results on a realistic BT system with comparisons in different domains are reported.  {\color{black}For comparisons, in the nominal case,  a multi-rate MPC in~\cite{zhang2018}, two single-layer MPCs described in~\cite{rawlings2017model}, a traditional decentralized PID controller, and a sliding mode controller (SMC) in~\cite{2020Performance} are used for comparison; while  two single-layer robust MPC regulators designed according to~\cite{kiaei2017tube} are also adopted in the perturbed case, two additional single-layer nonlinear MPC regulators in~\cite{lu2010study} and the sliding mode controller in~\cite{2020Performance} are introduced when applied to the nonlinear systems.}
\subsection{Description of the BT system and its linearized model}
A $160$ {\color{black}MW} BT system in~\cite{aastrom1987dynamic} is considered and its dynamic diagram is presented in Figure~\ref{fig:boiler-turbine}. The input variables applied to the boiler are the fuel flow $q_f$ ($\rm kg/s$) and feedwater flow $q_w$ ($\rm kg/s$), while the controlled variables of the boiler are drum pressure $P$ ($\text{ kg/cm}^2$) and water level $L$ ($\text{m}$). The control and controlled variables of the turbine are the steam control $q_s$ ($\rm kg/s$) and the electrical power output $Q$ {\color{black}($\text{MW}$)}. Typically, the goal of BT control is to regulate the electrical power to meet the load demand profile meanwhile to minimize the variations of internal variables such as water level and drum pressure  within their safe sets. Moreover, drum pressure must also be controlled properly in the operation range to respond to possible turbine speed changes caused by load demand variations. {\color{black}Many works have been addressed at this point that focus on deriving satisfactory closed-loop control performance of electrical power plants, see e.g.~\cite{chen2013multi,liu2009nonlinear,liu2013nonlinear,liu2018economic,liu2019fuzzyeconomic}}. 
In this scenario, the control related to the output variables such as electrical power and drum pressure is a major issue that must be tackled properly to respond to frequent load demand variations, while the water level can be adjusted smoothly under its constraint with the possibility to follow its desired value. This makes it reasonable in this case to apply the proposed dual-level control algorithms.
{\color{black}We adopt the  nonlinear dynamic model of a BT system described in \cite{aastrom1987dynamic} as the case study. In the nonlinear system, the state variables are $\rho$, $P$, $Q$, where $\rho$ is the fluid density $(\rm kg/cm^3)$ that establishes a static mapping with the water level. The control variables are limited by  $0\leq q_f,q_w,q_s\leq 1$ and their rate constraints are also considered, i.e., $-0.007\leq\dot q_f\leq 0.007$, $-2\leq \dot q_s\leq 0.2$, $-0.05\leq \dot q_w\leq 0.05$. The linearized model at an operation point $(\rho_r,\,P_r,\, Q_r)=(513.6,\break\,129.6,\,105.8)$, $(q_{w,r},\,q_{f,r},\, q_{s,r})=(0.663,\,0.505,\,0.828)$ is considered as the controlled system, i.e.,
	\begin{equation}\label{Eqn:CL_C}
	\left\{\begin{array}{l}
	\dot x=Ax+Bu\\
	[0.2cm]y=Cx,
	\end{array}\right. \qquad
	\end{equation}
	where
	$C=I$}, the state and output variables are $y=x=( \rho-\rho_r,\,P-P_r,\, Q-Q_r)$, while the input variables are $u=(q_w-q_{w,r},\, q_{f}-q_{f,r},\, q_s-q_{s,r})$.
The unitary step response of \eqref{Eqn:CL_C} is presented in Figure~\ref{fig:boiler-turbine_step}, which displays that the system outputs are strongly coupled, and the dynamics is not strictly  separable.
\begin{figure}[h!]
	\center
	\includegraphics[width=0.7\columnwidth]{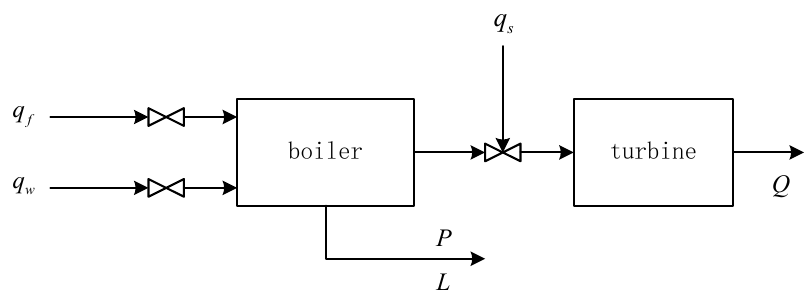}
	\caption{Diagram of the BT dynamics.}
	\label{fig:boiler-turbine}
\end{figure}
\begin{figure}[h!]
	\center
	\includegraphics[width=0.7\columnwidth]{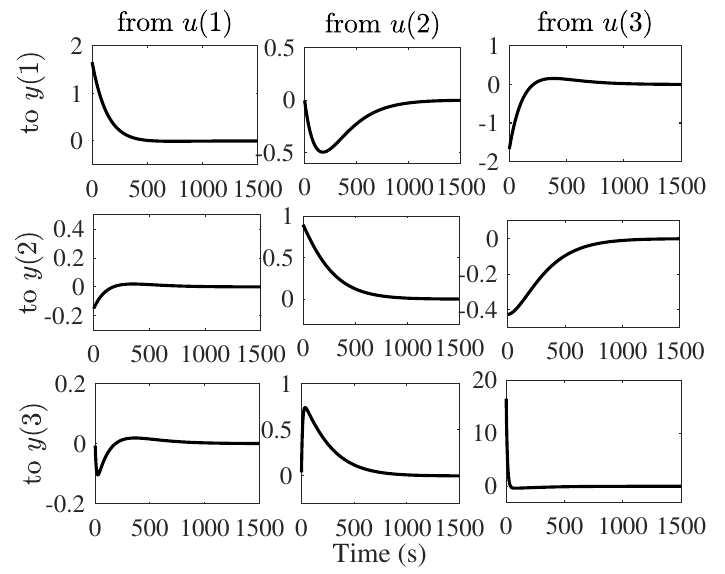}
	\caption{Unitary impulse response of the BT dynamics.}
	\label{fig:boiler-turbine_step}
\end{figure}
%
\subsection{Devising the D-MPC and Incremental D-MPC regulators}
In order to implement the proposed dual-level control algorithms, the system's continuous-time model \eqref{Eqn:CL_C} has been sampled with
$\Delta t=1$s to obtain the discrete-time counterpart in the fast time scale. The resulting system has been rewritten to derive model\break~\eqref{Eqn:sigma_s_f}, where the input, state, and output variables associated with $\Sigma_s$ to be controlled smoothly are chosen as $u_s=q_w-q_{w,r}$, $x_s=\rho-\rho_r$, and $y_s=x_s$ while the corresponding ones associated with $\Sigma_f$ to be controlled in a prompt fashion are $u_f=(q_{f}-q_{f,r},\, q_s-q_{s,r})$, $x_f=(P-P_r,\, Q-Q_r)$, and $y_f=x_f$.
The resulting model has been re-sampled with $N=20$ to obtain \eqref{Eqn:CL_k}  and~\eqref{Eqn:sigma_s_f-k-im} to be used at the high level.\\
\subsubsection{Design of the D-MPC regulator}
\begin{itemize}
	\item The high-level MPC \eqref{Eqn:HLoptimiz} with cost \eqref{Eqn:HL_cost} has been implemented with  $Q_{\rm\scriptscriptstyle H}=I$ and $R_{\rm\scriptscriptstyle H}=\text{diag}(2,\ 20,\ 20)$, and prediction horizon $N_{\rm\scriptscriptstyle H}=20$. The control gain matrix $K_{\rm\scriptscriptstyle H}$ is selected by solving an infinite horizon LQ problem. The terminal penalty $P_{\rm\scriptscriptstyle H}$ is computed according to \eqref{Eqn:lyap}.  
The terminal set has been chosen according to the algorithm described in \cite{hu2008model}.
	\item The low-level shrinking horizon MPC \eqref{Eqn:LLoptimiz} with cost \eqref{Eqn:LLoptimiz_cost} has been designed with $Q=I$ and $R=\text{diag}(1,\, 1,\, 10)$.
\end{itemize}
\subsubsection{Design of the Incremental D-MPC regulator}
\begin{itemize}
	\item The high-level MPC \eqref{Eqn:HLoptimiz-im} with cost \eqref{Eqn:HL_cost-im} has been implemented with $N_{\alpha}=2$ (see Algorithm~\ref{Atm:2}), $\bar Q_{\rm\scriptscriptstyle H}=I$ and $\bar R_{\rm\scriptscriptstyle H}=\text{diag}(2,\, 20,\, 20)$, and prediction horizon $\bar N_{\rm\scriptscriptstyle H}=20$. The control gain matrix $\bar K_{\rm\scriptscriptstyle S, H}$ is selected by solving the corresponding infinite horizon LQ problem. The terminal penalty is computed according to \eqref{Eqn:lyap}.
Likewise, the terminal set has been chosen according to the algorithm described in \cite{hu2008model}.
	\item The low-level shrinking horizon MPC \eqref{Eqn:LLoptimiz_rev} with cost \eqref{Eqn:LLoptimiz_cost_rev} has been designed with $\bar Q=I$ and $R=\text{diag}(1,\, 1,\, 10)$.
\end{itemize}
{\color{black}
\subsection{Simulation results: control of the linear nominal model}
	The proposed dual-level control algorithms have been applied to the linear BT system by solving an output reference tracking problem. The output set-point $y_r=(10,2,-2)$ is initially considered; while at time $t=400$$s$, the reference value has been reset according to the new load profile, i.e.,  $y_r=(5,1,4)$. The dual-level control algorithms have been implemented from null initial conditions. In the following, the simulation results have been reported including the comparisons with a multi-rate MPC in~\cite{zhang2018}, two single-layer MPC regulators designed according to~\cite{rawlings2017model} working at different time scales, and a traditional decentralized PID controller. 
\subsubsection{Design of the multirate MPC~\cite{zhang2018}}
 Note that, due to the usage of finite impulse response representation, the model used for the multirate MPC must be strictly stable.  However, the considered system in this paper has a pole on the unitary circle. In order to implement the multirate MPC successfully,  a feedback compensator $u_s=ky_s+v$ has been used firstly, where $v$ is defined as an auxiliary control variable, and the feedback gain $k$ is chosen as $k=-0.005$. For fair comparison, the design parameters $Q_s$ and $R_s$ have been selected coincident with the proposed MPC algorithms, i.e., $Q_s=\text{diag}\{1,2,\cdots,2,20,\cdots,20\}$, $R_s=2$.
\subsubsection{Design of the single-layer MPCs~\cite{rawlings2017model}}
Two single-layer stabilizing MPCs are designed working in slow and fast time scales respectively. The sampling time period of the slow and fast single-layer MPCs have been chosen as $\Delta t=20s,\,1s$, and the prediction horizons of the slow and fast MPC has been set as $N_{\rm\scriptscriptstyle H}$ and $N$ respectively.  The design parameters $Q$ and $R$ have been set to the values same as the high-level of the D-MPC.
\subsubsection{Design of the decentralized PID controller}
The decentralized PIDs, one for each input/output pair, have been designed with all the selected tuning parameters listed in Table \ref{tab:Tab_com0}.
\begin{table}[h!tb]
	\centering \caption{Tuning parameters of the decentralized PIDs}
	\label{tab:Tab_com0}
	\vskip 0.2cm
	\scalebox{0.8}{
		\begin{tabular}{ccccccc}
			\toprule[1pt]
			\multicolumn{3}{c}{Control pair}  & Proportional ($\text P$) &  Integral (I)&Derivative (D)\\
			\midrule[1pt]
			\multicolumn{3}{c}{$(u_s,y_s)$} & 0.019 &2$\cdot10^{-4}$&-0.07\\
			\hline
			\multicolumn{3}{c}{$(u_f(1),y_f(1))$} & 0.24 &0.006&-1\\
			\hline
			\multicolumn{3}{c}{$(u_f(2),y_f(2))$} & -0.035 &-4.6$\cdot10^{-4}$ &0.36\\
			\bottomrule[1pt]
		\end{tabular}
	}
\end{table}
\subsubsection{Simulation results}
	\begin{figure}[h!]
	\center
	\includegraphics[width=0.9\columnwidth]{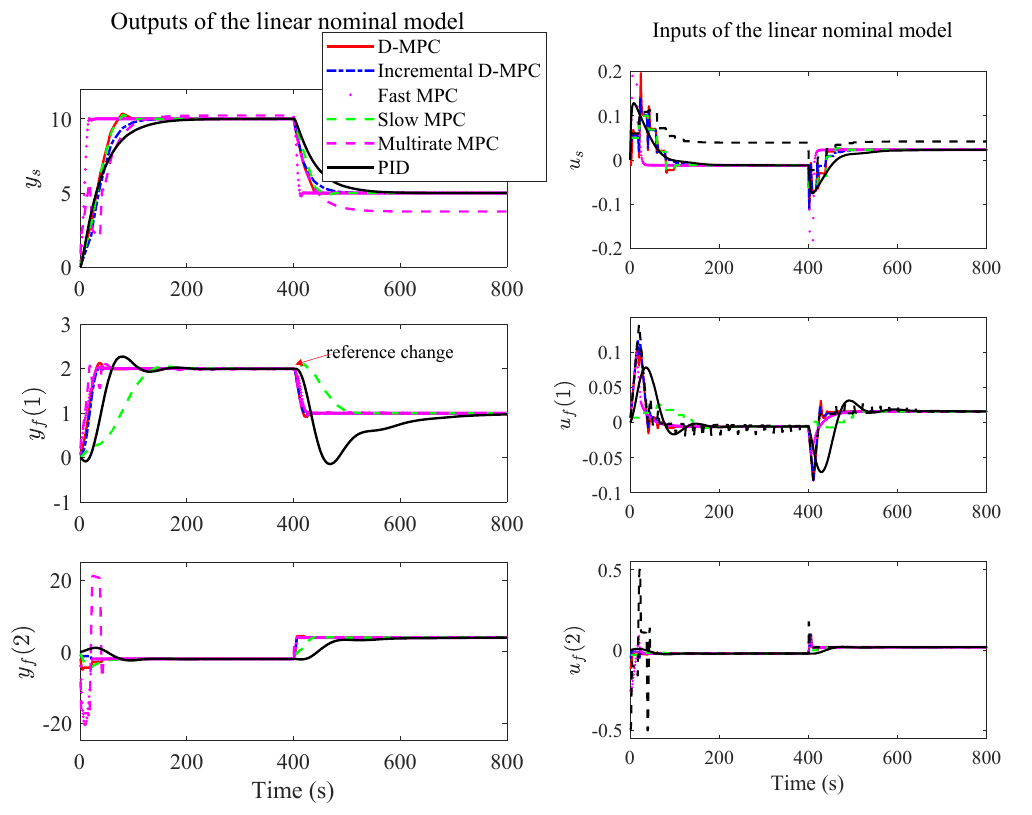}
	\caption{{\color{black}Output and control variables of the controlled linear model.}
	}
	\label{fig:nominal-comparison}
\end{figure}
\begin{figure}[h!]
	\center
	\includegraphics[width=0.8\columnwidth]{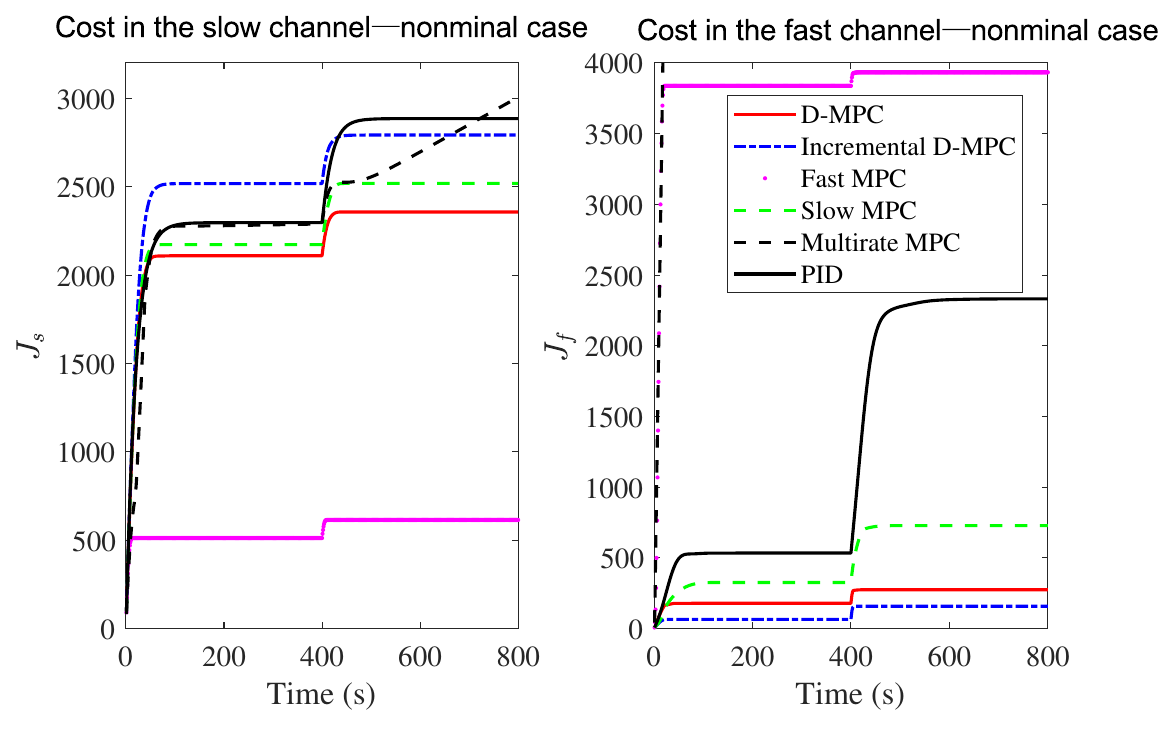}
	\caption{{\color{black}The comparison of variations of cumulative square tracking errors among all the controllers in the nominal case: The proposed D-MPC and \emph{Incremental} D-MPC have gained the smaller tracking costs of the fast (crucial) output $y_f$ at the expense of larger ones of the slow (less crucial) output $y_s$.}
	}
	\label{fig:cost_nominal}
\end{figure}
	All the comparative
	simulation experiments have been implemented within
	Yalmip toolbox installed in MATLAB environment, see~\cite{Lofberg2004}, in a Laptop with Intel Core i9-9880H 2.30 GHz running Windows 10 operating
	system. The simulation results have been reported in Figures~\ref{fig:nominal-comparison}-\ref{fig:cost_nominal} and Figure~\ref{fig:com_time}. It can be seen in Figure~\ref{fig:nominal-comparison} that, after an initial transient, inputs and outputs return to their nominal values, until the change of the reference occurs when the dual-level MPCs and single-layer MPCs properly react to bring the input and output variables to their new steady-state values, while the multirate MPC and decentralized PIDs react more slowly to reference variations. Note that, the proposed dual-level algorithms exhibit better control performances than other approaches for the crucial pair $(u_f,y_f)$, while the fast MPC performs the best for the pair $(u_s,y_s)$, which in fact can be controlled smoothly in the considered problem. 
		In Figure~\ref{fig:cost_nominal}, the cumulative square tracking errors $J_s=\sum_{i=1}^{N_{\rm sim}}\|y_s(i)-y_{s,r}\|^2$, and $J_f=\sum_{i=1}^{N_{\rm sim}}\|y_f(i)-y_{f,r}\|^2$, along the simulation steps from $N_{\rm sim}=1$ to $800$ are displayed for all the approaches, which show that, the proposed algorithms result smaller values of cost $J_f$ than other approaches at the expense of a larger cost on $J_s$. The cost $J_s$ with the fast MPC is the lowest but at the expense of a larger cost on $J_f$, i.e., a degradation of the control performance on $(u_f,y_f)$.
		 Also, the cost $J_f$ with the \emph{Incremental} D-MPC is smaller than that with the D-MPC at the price of a slightly larger $J_s$.
		This reveals that the proposed  D-MPC and \emph{Incremental} D-MPC show strong points in imposing separable closed-loop dynamics, i.e., controlling the pair $(u_f,y_f)$ promptly while regulating the less crucial pair $(u_s,y_s)$ in a smoother fashion. Also, the \emph{Incremental} D-MPC outperforms the D-MPC in this respect. As for computational resources, the average computational times of the proposed algorithms are smaller than that of the fast MPC in the nominal case, see Figure~\ref{fig:com_time}.

\begin{figure}[h!]
	\center
	\includegraphics[width=0.6\columnwidth]{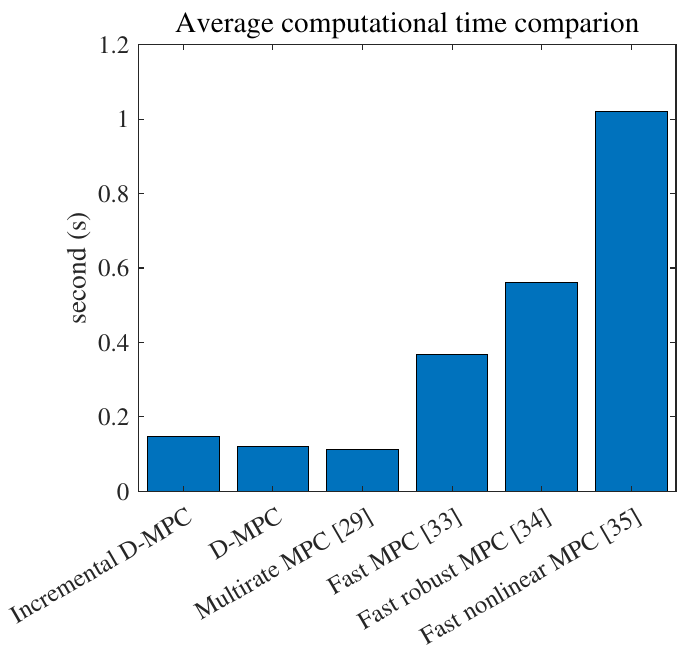}
	\caption{{\color{black}Computational times comparison. The computational times of the proposed algorithms are smaller than that of the fast single-layer MPCs.}
	}
	\label{fig:com_time}
\end{figure}
	\subsection{Simulation results: control of the linear model with uncertainties}
			
	\begin{figure}[h!]
		\center
		\includegraphics[width=0.6\columnwidth]{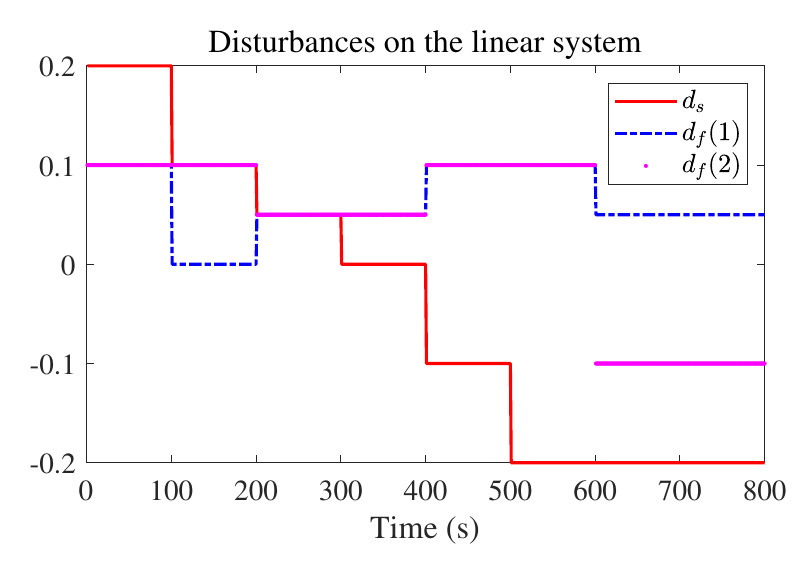}
		\caption{{\color{black}Disturbances on the linear model.}
		}
		\label{fig:disturbance}
	\end{figure}	
	\begin{figure}[h!]
		\center
		\includegraphics[width=0.9\columnwidth]{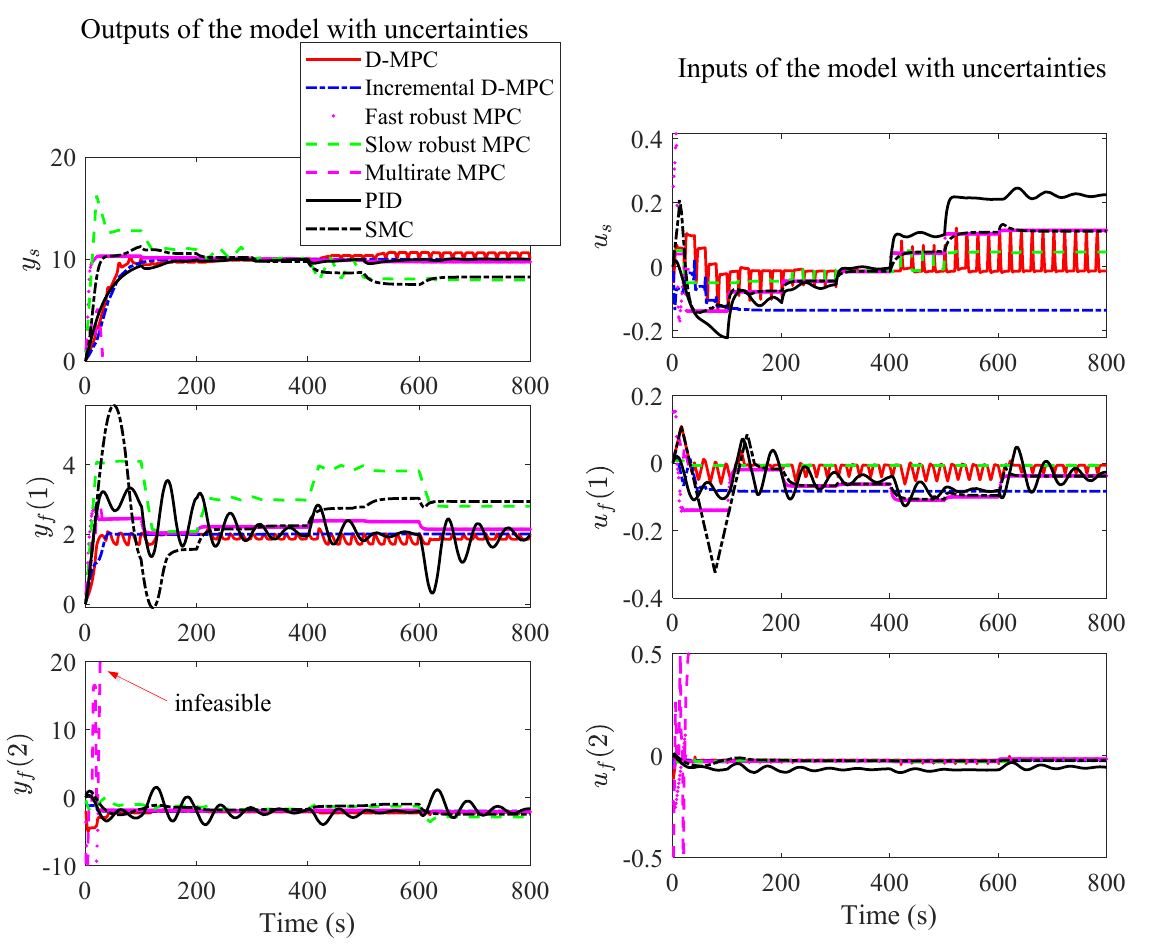}
		\caption{{\color{black}Output and control variables of the controlled linear model with uncertainties.} 
		}
		\label{fig:nominal-comparison-dis}
	\end{figure}
	\begin{figure}[h!]
		\center
		\includegraphics[width=0.8\columnwidth]{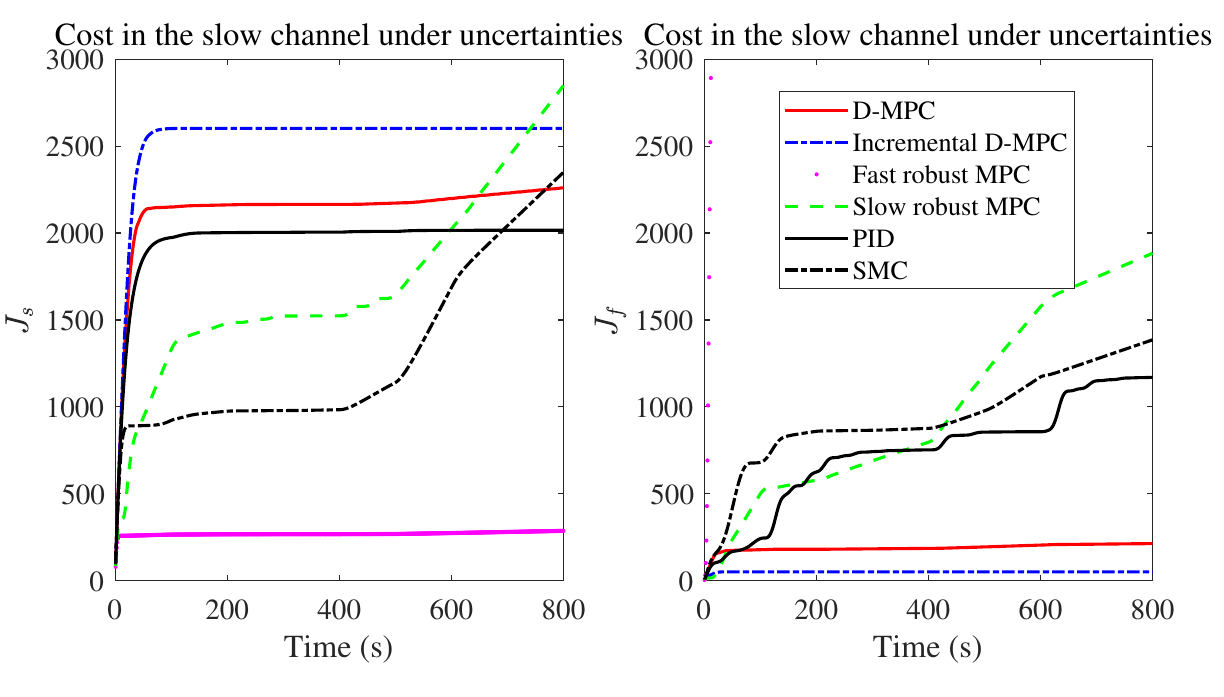}
		\caption{{\color{black}The comparison of variations of cumulative square tracking errors among different controllers in the perturbed case: The proposed \emph{Incremental} D-MPC is the only one that has realized offset-free control for both the fast and slow control channel. Note that, the \emph{Incremental} D-MPC has gained the smallest tracking cost of the fast (crucial) output $y_f$ at the expense of the largest one of the slow (less crucial) output $y_s$.}  
		}
		\label{fig:cost_robust}
	\end{figure}
	To further verify the capability of the proposed algorithms in dealing with disturbances. A bounded unknown step-wise disturbance, i.e., $(-0.2,0.05,-0.1)\leq d\leq(0.2,0.1,0.1) $, has been added to the controlled system, see~Figure~\ref{fig:disturbance}. In the proposed controller, the control constraints have been properly tightened according to~\eqref{Eqn:fea-con-dmpc} and the terminal constraint has been computed according to~\cite{marruedo2002input}. For comparison, the multirate MPC~\cite{zhang2018},  two single-layer robust MPC regulators in~\cite{kiaei2017tube}, the decentralized PID controller, and the sliding mode controller in~\cite{2020Performance} have been used.  In the robust MPCs, the design parameters have been chosen similar to the nominal MPCs, except that the control constraints have been tightened according to the robust invariant set for real constraint satisfaction under perturbations and the optimization on the initial nominal state has been considered, see~\cite{kiaei2017tube}. In the sliding mode controller, all the parameters are fine tuned according to the design procedures in~\cite{2020Performance}. In the simulation tests, the output set-point regulation with $y_r=(10,2,-2)$ has been  considered. The corresponding simulation results are presented in Figures~\ref{fig:nominal-comparison-dis}-\ref{fig:cost_robust}. The results show that, the proposed \emph{Incremental} D-MPC can realize offset-free control for all the outputs, which is not yet realized by the D-MPC, the multirate MPC, the robust MPCs, the PIDs, and the SMC, since most of the cumulative costs are still increasing at the terminal simulation time, see Figure~\ref{fig:cost_robust}. Also, the multirate MPC is early terminated due to the in-feasibility issue caused by disturbances. It reveals that,  the proposed \emph{Incremental} D-MPC outperforms  the multirate MPC, robust MPCs, and decentralized PIDs, and the SMC,  especially in enforcing satisfactory control performance for the pair $(u_f,y_f)$, see again Figure~\ref{fig:cost_robust}. Also, the average computational times with the proposed algorithms are smaller than that of the fast robust MPC, see Figure~\ref{fig:com_time}.}
 {\color{black}
 		\begin{figure}[h!]
 	\center
 	\includegraphics[width=0.9\columnwidth]{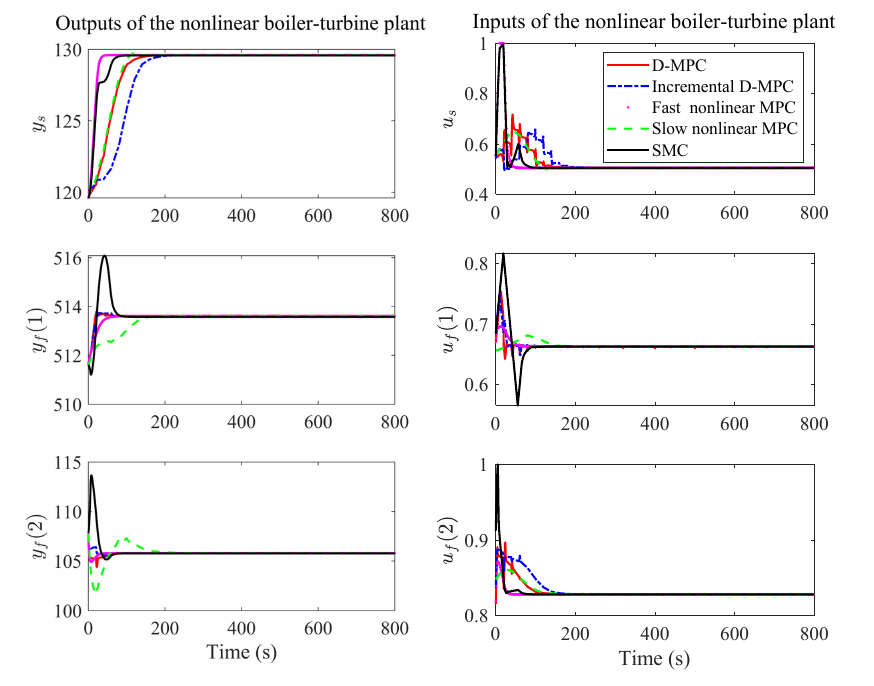}
 	\caption{{\color{black}Output and control variables of the nonlinear controlled boiler-turbine plant.}
 	}
 	\label{fig:nominal-comparison-non}
 \end{figure}
 \begin{figure}[h!]
 	\center
 	\includegraphics[width=0.8\columnwidth]{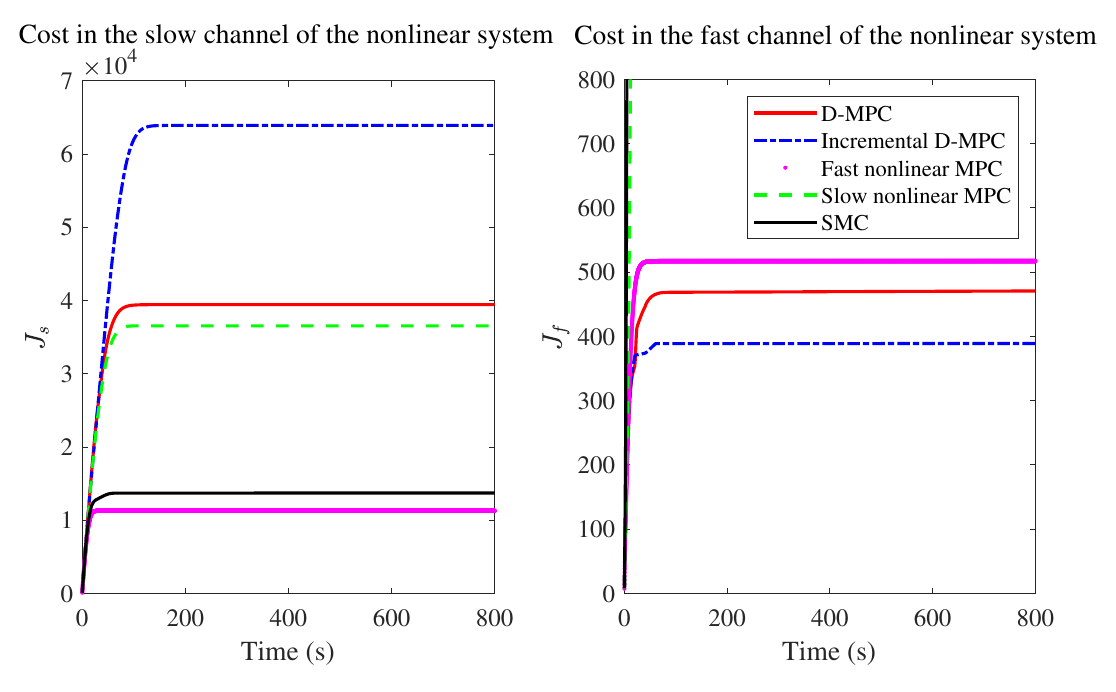}
 	\caption{{\color{black}The comparison of variations of cumulative square tracking errors of the nonlinear controlled systems: The proposed algorithms exhibit better control performance for the fast output $y_f$ than the nonlinear MPCs. Note that, the proposed \emph{Incremental} D-MPC has gained the smallest tracking cost of the fast (crucial) output $y_f$ at the expense of the largest one of the slow (less crucial) output $y_s$.}  
 	}
 	\label{fig:cost_non}
 \end{figure}
\subsection{Simulation results: application to the nonlinear system}
Besides the verification of the developed theory in the nominal and perturbed cases, we have also applied the proposed algorithms to the original nonlinear systems, see \cite{aastrom1987dynamic}. Two single-layer nonlinear MPC regulators designed according to~\cite{lu2010study} and the SMC in~\cite{2020Performance} are used for fair comparison. In the nonlinear MPCs, the prediction horizons have been chosen as 20. The performance index adopted has been $J=\sum_{j=0}^{N_p-1}\|x(t+j)-x_r\|^2_{Q_{\rm\scriptscriptstyle H}}+\|u(t+j)-u_r\|^2_{R_{\rm\scriptscriptstyle H}}+\|x(t+N_p)-x_r\|_P^2$, where the terminal cost is added to guarantee stability and $P$ is computed using the nominal linearized model.
In the simulation test, the control goal has been to drive the state variable to the reference $x_r=(513.6,\break\,129.6,\,105.8)$ from the initial condition $x_0=x_r+(10,2,-2)$. 
	 The  simulation results are reported in Figures~\ref{fig:nominal-comparison-non}-\ref{fig:cost_non}. The results show that, the proposed \emph{Incremental} D-MPC achieves the best control performance for $(u_f,y_f)$ at the expense of the control performance degradation on $(u_s,y_s)$, see Figure~\ref{fig:cost_non}. The fast nonlinear MPC and SMC perform better for $(u_s,y_s)$ but worse than the proposed D-MPC and \emph{Incremental} D-MPC in terms of that for $(u_f,y_f)$, hence it does not fulfill the control goal specified in this paper. 
	 Also, as shown in Figure~\ref{fig:com_time}, the average computational times of the proposed algorithms are much smaller than that of the fast nonlinear MPC. 
}
{\color{black}
\subsection{Discussions}
One can conclude from the above simulation tests that, the proposed algorithms, especially \emph{Incremental} D-MPC, have fulfilled the control objective considered in this paper, i.e., generating satisfactory fast dynamics for the fast (crucial) pair $(u_f,y_f)$ and smooth dynamics for the slow (less crucial) pair $(u_s,y_s)$, see Figures~\ref{fig:cost_nominal},~\ref{fig:cost_robust}, and~\ref{fig:cost_non}. The former is enforced at the expense of the performance degradation of $(u_s,y_s)$. However, the control objective is not well met by the single-layer fast MPCs working in the basic time scale, see again Figures~\ref{fig:cost_nominal},~\ref{fig:cost_robust}, and~\ref{fig:cost_non}, where differently, the control performances of the less crucial $(u_s,y_s)$ are the best ones among all the controllers in the simulation tests. Note that, the proposed algorithms might not outperform single-layer MPCs when the overall control performance is a major concern, which is not in the scope of this paper. The computational times of the proposed algorithms are much smaller than that of the fast MPCs, due to the shrinking horizon strategy used in the basic time, see Figure~\ref{fig:com_time}. 
 
Different from output-feedback controllers, the proposed approaches do not rely on an observer for estimating disturbances. The simulation results in Figures~\ref{fig:nominal-comparison-dis}-\ref{fig:cost_robust} reveals that the proposed approaches are robust to time-varying disturbances, and the \emph{Incremental} D-MPC can realize offset-free control under unknown time-varying piece-wise disturbances. One limitation of the proposed algorithms lies in the boundedness assumption of the disturbances for the recursive feasibility guarantee of constrained control problems.
}
\section{Conclusions}
In this paper, two dual-level MPC control algorithms have been proposed for
linear multi-timescale systems with input constraint.  The proposed MPC algorithms rely on clear time separation, so allow to deal with control problems in different channels. In view of their main properties, the proposed algorithms are, based on the solution based on MPC with dual-level structure, suitable not only to cope with control of singularly perturbed systems but
also to impose different closed-loop dynamical performance for systems with non-separable open-loop dynamics.
 
 {\color{black}The recursive feasibility and convergence of the proposed D-MPC and \emph{Incremental} D-MPC are proven under suitable assumptions.  The effectiveness of the proposed algorithms is tested rigorously in different simulation scenarios, including numerous comparisons with different classic controllers. The simulation results show that both the proposed D-MPC and \emph{Incremental} D-MPC are effective in imposing closed-loop separable dynamics and can deal with unknown bounded time-varying disturbances. Also, the latter can obtain offset-free control under unknown and rapidly varying step-wise disturbances without using a disturbance estimator.} 
 
Future work will consider extending the proposed framework to solving multi-rate control problems for systems with large time scales, {\color{black}possibly relies on a cloud-edge computing structure; and apply the proposed approaches to multi-agent control systems.}
%
\section{Acknowledgment}
The first author wishes to thank Prof. Riccardo Scattolini and Prof. Marcello Farina from Politecnico di Milano, for their fruitful discussions.
%
%
%
%
\appendices
\section{}
 
{\color{black}
\subsection{Proof of Proposition~\ref{prop:obser-con}}\label{sec:prop-con-proof}
According to PBH detectability rank test, the pair $(A,\,C)$ is detectable if and only if $\text{rank}(\begin{bmatrix}
\lambda I-A&C
\end{bmatrix})=n$, $\forall\,\lambda \in \mathbb{C}$ and $|\lambda| \geq 1$. An equivalent form to this condition is that $v=0$ is the unique solution to the following linear equations
\begin{equation}\label{Eqn:con-cond}
\left\{ \begin{array}{c}
Av=\lambda v\\
Cv=0,
\end{array}\right.
\end{equation}
 $\forall\,\lambda\in \mathbb{C}$ and $|\lambda|\geq 1$.
From~\eqref{Eqn:con-cond}, $v=0$ is the unique solution to $\lambda^{i-1}Av=\lambda^i v,\,
Cv=0,\,\forall\, i\in\mathbb{N}_+$, which is $A^iv=\lambda^iv,\, Cv=0,\,\forall\, i\in\mathbb{N}_+$.
In view of this, recalling that $(A,\,C)$ is detectable,  it holds that $v=0$ is the only solution to
	\begin{equation*}
\left\{	\begin{array}{c}
	A^{\scriptscriptstyle[N]}v=\mu v\\
	Cv=0,
	\end{array}\right.
	\end{equation*}
where $\mu=\lambda^{N}$, which implies  $(A^{\rm \scriptscriptstyle[N]},\, C)$ is observable for all the modes that their poles $|\lambda|\geq 1$. Hence,  Proposition~\ref{prop:obser-con} holds. }\hfill $\square$

\subsection{Proof of Theorem~\ref{theorem}}\label{sec:theo-proof}
 
\subsubsection{Recursive feasibility of the D-MPC (i.e., high-level problem~\eqref{Eqn:HLoptimiz} and low-level problem \eqref{Eqn:LLoptimiz})}
As the problem \eqref{Eqn:HLoptimiz} is assumed to be feasible at time $k=0$, with resorting to Mathematical Induction technique, one can prove the closed-loop recursive feasibility by verifying that if \eqref{Eqn:HLoptimiz} is feasible at any time $k$, then
\begin{itemize}
	\item [\emph{(i)}]  the low-level problem \eqref{Eqn:LLoptimiz} is feasible at any fast time $h\in[kN,kN+N)$;
	\item [\emph{(ii)}] also the high-level problem \eqref{Eqn:HLoptimiz} is feasible at the subsequent slow time instant $k+1$.
\end{itemize}
{\color{black}
First, we show that condition \emph{(i)} can be verified. To proceed, we assume that the high-level problem \eqref{Eqn:HLoptimiz} is feasible at time $k$. 
Assume $\delta u(kN),\cdots, \delta u(h|h), \cdots, \delta u(kN+N-1|h)\in \Delta {\mathcal{U}}\times \Delta \hat{\mathcal{U}}^{N-1}$ is a feasible solution at time $h\in[kN,kN+N-2]$ and the terminal state constraint is verified. Letting $h=kN+j$, one has
\begin{equation}\label{Eqn:current-terminal-dmpc}
\begin{array}{ll}
\hat x^{\scriptscriptstyle[N]}(k+1|k)=&
\sum_{i=0}^{N-1}A^{i}B u(kN+i|h)+A^{N}x(kN)+\\&\sum_{i=0}^{j-1}A^{N-i-1}d(kN+i)
\end{array}
\end{equation}
One can also write the predicted terminal state at time $h+1$, i.e.,
\begin{equation}\label{Eqn:next-terminal-dmpc}
\begin{array}{ll}
\hat x(kN+N|h+1)=\hspace{-3mm}&\sum_{i=0}^{N-1}A^{i}B u(kN+i|h+1)+A^{N}x(kN)+\\&\sum_{i=0}^{j}A^{N-i-1}d(kN+i)
\end{array}
\end{equation}
To ensure the recursive feasibility, it is required to enforce $\hat x(kN+N|h+1)=\hat x^{\scriptscriptstyle[N]}(k+1|k)$. By difference of~\eqref{Eqn:next-terminal-dmpc} and \eqref{Eqn:current-terminal-dmpc}, leads to
 
\begin{equation}\label{Eqn:con_delta}
\begin{array}{lll}
\sum_{i=j+1}^{N-1}A^{N-i-1}B u(kN+i|h+1)=-A^{N-j-1}d(h)+\\
\sum_{i=j+1}^{N-1}A^{N-i-1}B u(kN+i|h)\in \Delta L^j {\mathcal{U}}\oplus L^{j+1} \hat{\mathcal{U}}^{j+1},
\end{array}
\end{equation}
in view of condition~\eqref{Eqn:fea-con-dmpc}. Hence, the recursive feasibility at the low level follows.
 
As for \emph{(ii)}, first note that, one can compute the gap between the real state and the predicted one, i.e.,
\begin{equation}\label{Eqn:disturbance-real-dmpc}
{x}^{\scriptscriptstyle[N]}(k+1)-{\hat x}^{\scriptscriptstyle[N]}(k+1|k)=\tilde d^{\scriptscriptstyle[N]}(k)
\end{equation}
Note that, ${\hat x}(kN+N|kN+N-1)= \hat {x}^{\scriptscriptstyle[N]}(k+1)$ and $x(kN+N)-{\hat x}(kN+N|kN+N-1)=d(kN+N-1)$. One promptly has $$\tilde d^{\scriptscriptstyle[N]}(k)=d(kN+N-1).$$
 Hence, one can also compute:
\begin{equation}\label{Eqn:disturbance-real-hat-dmpc}
{\hat x}^{\scriptscriptstyle[N]}(k+ N_{\rm\scriptscriptstyle H}|k)-{\hat x}^{\scriptscriptstyle[N]}(k+ N_{\rm\scriptscriptstyle H}|k+1)= ( A^{\scriptscriptstyle[N]})^{ N_{\rm\scriptscriptstyle H}-1}d(kN+N-1)
\end{equation}
Assume that at any slow time instant $k$ the optimal control sequence of \eqref{Eqn:HLoptimiz} can be found, i.e., $\overrightarrow {u^{\scriptscriptstyle[N],o}}(k:k+N_{\rm\scriptscriptstyle H}-1|k)=\big(u^{\scriptscriptstyle[N],o}(k|k),\cdots,u^{\scriptscriptstyle[N],o}(k+N_{\rm\scriptscriptstyle H}-1|k)\big)$ such that $x^{\scriptscriptstyle[N]}(k+N_{\rm\scriptscriptstyle H}|k)\in \mathcal{X}_{\rm\scriptscriptstyle F}^s$.
 Let the input sequence  $\overrightarrow {u^{\scriptscriptstyle[N],s}}(k+1:k+N_{\rm\scriptscriptstyle H}|k+1)=\big(u^{\scriptscriptstyle[N],o}(k+1|k),\cdots,u^{\scriptscriptstyle[N],o}(k+N_{\rm\scriptscriptstyle H}-1|k),K_{\rm\scriptscriptstyle H}(x^{\scriptscriptstyle[N]}(k+N_{\rm\scriptscriptstyle H}|k)-x_r)+u_r\big)$ be a candidate choice at the next time instant $k+1$. As condition~\eqref{Eqn:state-terminal-size-dmpc} is assumed, it holds that $ {\hat x}(k+ N_{\rm\scriptscriptstyle H}|k+1)\in {\mathcal{X}}_{\rm\scriptscriptstyle F}$. In view of this, $ {\hat x}(k+ N_{\rm\scriptscriptstyle H}+1|k+1)\in {\mathcal{X}}_{\rm\scriptscriptstyle F}^s$ can be verified in view of the definitions of $K_{\rm\scriptscriptstyle H}$ and ${\mathcal{X}}_{\rm\scriptscriptstyle F}^s$.  Hence, the recursive feasibility of \eqref{Eqn:HLoptimiz} follows.
}
\subsubsection{Convergence of the closed-loop system}
 
We first prove the convergence of the high-level problem~\eqref{Eqn:HLoptimiz}. As the disturbance $d=0$, it holds that $\hat x=x$, $\hat y=y$.
Denote by  $J_{\rm\scriptscriptstyle H}^o(x^{\scriptscriptstyle[N]}(k))$ the optimal cost associated with $\overrightarrow {u^{\scriptscriptstyle[N],o}}(k:k+N_{\rm\scriptscriptstyle H}-1|k)$ at time $k$ and by $J_{\rm\scriptscriptstyle H}^s(x^{\scriptscriptstyle[N]}(k+1|k))$ the suboptimal cost associated with $\overrightarrow {u^{\scriptscriptstyle[N],s}}(k+1:k+N_{\rm\scriptscriptstyle H}|k+1)$ at time $k+1$.
It is possible to write
\begin{equation}\label{Eqn:cost_decre1}
\begin{array}{lll}
&J_{\rm\scriptscriptstyle H}^s(x^{\scriptscriptstyle[N]}(k+1|k))-J_{\rm\scriptscriptstyle H}^o(x^{\scriptscriptstyle[N]}(k))=\\&=-(\|y^{\scriptscriptstyle[N]}(k)-y_r\|^2_{Q_{\rm\scriptscriptstyle H}}+\|u^{\scriptscriptstyle[N],o}(k|k)-u_r\|^2_{R_{\rm\scriptscriptstyle H}})+\\
&\ \  \ \|y^{\scriptscriptstyle[N]}(k+N|k)-y_r\|_{Q_{\rm\scriptscriptstyle H}}^2+ \|K_{\rm\scriptscriptstyle H}(x^{\scriptscriptstyle[N]}(k+N_{\rm\scriptscriptstyle H}|k)-x_r)\|_{R_{\rm\scriptscriptstyle H}}^2+\\
&\ \ \ \|F_{\rm\scriptscriptstyle H}(x^{\scriptscriptstyle[N]}(k+N|k)-x_r)\|^2_{P_{\rm\scriptscriptstyle H}}-\|x^{\scriptscriptstyle[N]}(k+N|k)-x_r\|^2_{P_{\rm\scriptscriptstyle H}}\\
&=-(\|y^{\scriptscriptstyle[N]}(k)-y_r\|^2_{Q_{\rm\scriptscriptstyle H}}+\|u^{\scriptscriptstyle[N],o}(k|k)-u_r\|^2_{R_{\rm\scriptscriptstyle H}})\\
&\ \ \ +\|x^{\scriptscriptstyle[N]}(k+N|k)-x_r\|^2_{	F_{\rm\scriptscriptstyle H}^{\top} P_{\rm\scriptscriptstyle H} F_{\rm\scriptscriptstyle H}-P_{\rm\scriptscriptstyle H}+C^{\top}Q_{\rm\scriptscriptstyle H}C+K_{\rm\scriptscriptstyle H}^{\top}R_{\rm\scriptscriptstyle H} K_{\rm\scriptscriptstyle H}=0}
\end{array}
\end{equation}
In view of \eqref{Eqn:lyap} and from \eqref{Eqn:cost_decre1}, one has
\begin{equation*}
\begin{array}{l}
J_{\rm\scriptscriptstyle H}^s(x^{\scriptscriptstyle[N]}(k+1|k))-J_{\rm\scriptscriptstyle H}^o(x^{\scriptscriptstyle[N]}(k))=\\
-(\|y^{\scriptscriptstyle[N]}(k)-y_r\|^2_{Q_{\rm\scriptscriptstyle H}}+\|u^{\scriptscriptstyle[N],o}(k|k)-u_r\|^2_{R_{\rm\scriptscriptstyle H}}).
\end{array}
\end{equation*}
Recalling the fact that $J_{\rm\scriptscriptstyle H}^o(x^{\scriptscriptstyle[N]}(k+1|k))\leq J_{\rm\scriptscriptstyle H}^s(x^{\scriptscriptstyle[N]}(k+1|k))$, then
\begin{equation}\label{Eqn:cost_monotonic1}
\begin{array}{l}
J_{\rm\scriptscriptstyle H}^o(x^{\scriptscriptstyle[N]}(k+1|k))-J_{\rm\scriptscriptstyle H}^o(x^{\scriptscriptstyle[N]}(k))\leq\\
-(\|y^{\scriptscriptstyle[N]}(k)-y_r\|^2_{Q_{\rm\scriptscriptstyle H}}+\|u^{\scriptscriptstyle[N],o}(k|k)-u_r\|^2_{R_{\rm\scriptscriptstyle H}}),
\end{array}
\end{equation}
which implies that
$J_{\rm\scriptscriptstyle H}^o(x^{\scriptscriptstyle[N]}(k+1|k))-J_{\rm\scriptscriptstyle H}^o(x^{\scriptscriptstyle[N]}(k))$ converges to zero. Moreover, from \eqref{Eqn:cost_monotonic1}, one has $J_{\rm\scriptscriptstyle H}^o(x^{\scriptscriptstyle[N]}(k))-J_{\rm\scriptscriptstyle H}^o(x^{\scriptscriptstyle[N]}(k+1|k))\geq\|y^{\scriptscriptstyle[N]}(k)-y_r\|^2_{Q_{\rm\scriptscriptstyle H}}+\|u^{\scriptscriptstyle[N],o}(k|k)-u_r\|^2_{R_{\rm\scriptscriptstyle H}}$, then $\|y^{\scriptscriptstyle[N]}(k)-y_r\|^2_{Q_{\rm\scriptscriptstyle H}}+\|u^{\scriptscriptstyle[N],o}(k|k)-u_r\|^2_{R_{\rm\scriptscriptstyle H}}\rightarrow 0$.  Recalling the definitions of $Q_{\rm\scriptscriptstyle H}$ and $R_{\rm\scriptscriptstyle H}$,  one has $\lim_{k\to+\infty} y^{\scriptscriptstyle[N]}(k)=y_r$  and $\lim_{k\to+\infty} u^{\scriptscriptstyle[N]}(k)=u_r$. In view of Proposition~\ref{prop:obser-con}, consequently $\lim_{k\to+\infty} x^{\scriptscriptstyle[N]}(k)=x_r$.
 
As for the convergence of the low-level problem~\eqref{Eqn:LLoptimiz}, assume that the high-level system variables have reached their reference values, i.e., $u^{\scriptscriptstyle[N]}(k)\equiv u_r$ $x^{\scriptscriptstyle[N]}(k)\equiv x_r$, $y^{\scriptscriptstyle[N]}(k)\equiv y_r$. Define $\delta x(k)=x(kN)-x_r$ and  $\delta y(k)=y(kN)-y_r$.
Along the same line in \cite{picasso2016hierarchical}, in view of  dynamics \eqref{Eqn:CL_final} at time instant $h=kN$, the low-level dynamics at the slow time scale is defined
\begin{equation}\label{Eqn:CL_delta_k}
\left\{\begin{array}{l}
\delta x(k+1)=A^N\delta x(k)+w(k)\\
[0.2cm]\delta y(k)=C\delta x(k),
\end{array}\right. \qquad
\end{equation}
where $w(k)=\sum_{j=0}^{N-1}A^{N-j-1}B\delta u(kN+j)$. Since $\delta x(k)=0,\, \forall k\geq 0$ (due to~\eqref{LLoptimiz_term_con}), it holds that $w(k)=0$. In view of the cost function at the low level,  the null sequence $\overrightarrow{\delta u}(h:(k+1)N-1)=0$ solves the problem \eqref{Eqn:HLoptimiz}, which implies that $\lim_{h\to+\infty} \delta u(h)=0$ and $\lim_{h\to+\infty} u(h)=u_r$. Finally, $\lim_{h\to+\infty} y(h)=y_r$ and\break $\lim_{h\to+\infty} x(h)=x_r$. \hfill $\square$
\subsection{Proof of proposition~\ref{prop1}}
According to PBH stabilizability rank test, the pair $(\bar A^{\scriptscriptstyle[N]},\,\bar B_s^{\scriptscriptstyle[N]})$ is stabilizable if and only if $\text{rank}(\begin{bmatrix}
\lambda I-\bar A^{\scriptscriptstyle[N]}&\bar B_s^{\scriptscriptstyle[N]}
\end{bmatrix})=n+p_s$, for $\lambda \in \mathbb{C}$ and $|\lambda| \geq 1$. An equivalent form to this condition is that $v=0$ is the unique solution to the following linear equations
\begin{equation}\label{Eqn:PBH_test}
\left\{\begin{array}{l}
(\bar A^{\scriptscriptstyle[N]})^{\top}v=\lambda v\\
[0.2cm] (\bar B_s^{\scriptscriptstyle[N]})^{\top}v=0,
\end{array}\right.\qquad
\end{equation}
where $\lambda\in \mathbb{C}$ and $|\lambda|\geq 1$.\\
In view of \eqref{Eqn:sigma_s-dus-k}, it is possible to write~\eqref{Eqn:PBH_test} in the form
\begin{equation}\label{Eqn:PBH_lambda1}
\begin{bmatrix}
I-\lambda I& 0\\\tilde C_s\tilde A^{\scriptscriptstyle[N]}&\tilde A^{\scriptscriptstyle[N]}-\lambda I\\
\tilde C_s\tilde B_s^{\scriptscriptstyle[N]}& \tilde B_s^{\scriptscriptstyle[N]}
\end{bmatrix}^{\top}v=0
\end{equation}
Since $\tilde A^{\scriptscriptstyle[N]}$ is stabilizable by Assumption~\ref{assum:tilde A-stable}, it is obvious to see that for $|\lambda|> 1$,  $v=0$ is the unique solution to~\eqref{Eqn:PBH_lambda1}. \\ For $\lambda=1$, $v=0$ is the unique solution to~\eqref{Eqn:PBH_lambda1} if and only if
$$\text{rank}\big(\begin{bmatrix}
\tilde C_s\tilde A^{\scriptscriptstyle[N]}&\tilde A^{\scriptscriptstyle[N]}-I\\\tilde C_s\tilde B_s^{\scriptscriptstyle[N]}&\tilde B_s^{\scriptscriptstyle[N]}
\end{bmatrix}^{\top}\big)=n+p_s.$$
As for $\lambda=-1$, $v=0$ is the unique solution to~\eqref{Eqn:PBH_lambda1} if and only if
$$\text{rank}\big(\begin{bmatrix}
2I&0\\
\tilde C_s\tilde A^{\scriptscriptstyle[N]}&\tilde A^{\scriptscriptstyle[N]}+I\\\tilde C_s\tilde B_s^{\scriptscriptstyle[N]}&\tilde B_s^{\scriptscriptstyle[N]}
\end{bmatrix}^{\top}\big)=n+p_s.$$
\hfill $\square$
\subsection{Proof of Theorem~\ref{theorem-im}}
 
\subsubsection{Recursive feasibility of the Incremental D-MPC (i.e., high-level problem~\eqref{Eqn:HLoptimiz-im} and low-level problem~\eqref{Eqn:LLoptimiz_rev})}
{\color{black}
As $N_{\alpha}$ is assumed to be reachable by Algorithm~\ref{Atm:2} such that \eqref{Eqn:HLoptimiz-im} is feasible at time $k=0$, along the same line of Section~\ref{sec:theo-proof}, we first prove the recursive feasibility for problem \eqref{Eqn:LLoptimiz_rev} in the fast time scale. To this end, we assume \eqref{Eqn:LLoptimiz_rev} is feasible at a time instant $h\in[kN,kN+N)$,  i.e., one can find the candidate control sequence $\Delta u(kN),\cdots, \Delta u(h|h), \cdots, \Delta u(kN+N-1|h)$ such that $\delta u(kN),\cdots, \delta u(h|h), \cdots, \delta u(kN+N-1|h)\in \Delta {\mathcal{U}}\times \Delta \hat{\mathcal{U}}^{N-1}$ and the terminal state constraint is verified. Along the same line with Appendix~\ref{sec:theo-proof}, one requires condition~\eqref{Eqn:con_delta}, 
 which is verified in view of~\eqref{Eqn:fea-con-dmpc}. Hence, the recursive feasibility at the low level follows.
 
As for a sketch of the proof of the feasibility at the high level, first note that, one can compute:
\begin{equation}\label{Eqn:disturbance-real}
\begin{array}{ll}
\Delta{x}^{\scriptscriptstyle[N]}(k+1)-\Delta{\hat x}^{\scriptscriptstyle[N]}(k+j|k)&={x}^{\scriptscriptstyle[N]}(k+1)-\hat {x}^{\scriptscriptstyle[N]}(k+1|k)\\
&=d(kN+N-1)
\end{array}
\end{equation}
 
Recalling that $\bar { x}^{\scriptscriptstyle[N]}=( y_s^{\scriptscriptstyle[N]},\ \Delta { x}^{\scriptscriptstyle[N]})$, one can also compute:
\begin{equation}\label{Eqn:disturbance-real-hat}
\bar{\hat x}^{\scriptscriptstyle[N]}(k+\bar N_{\rm\scriptscriptstyle H}|k)-\bar{\hat x}^{\scriptscriptstyle[N]}(k+\bar N_{\rm\scriptscriptstyle H}|k+1)=(\bar A^{\scriptscriptstyle[N]})^{\bar N_{\rm\scriptscriptstyle H}-1}Ed(kN+N-1)
\end{equation}
Let assume at time $k$ that the optimal control sequence \eqref{Eqn:HLoptimiz} is found, i.e., $\overrightarrow {\Delta u_s^{\scriptscriptstyle[N],o}}(k:k+\bar N_{\rm\scriptscriptstyle H}-1|k)=\big(\Delta u_s^{\scriptscriptstyle[N],o}(k|k),\cdots,\Delta u_s^{\scriptscriptstyle[N],o}(k+\bar N_{\rm\scriptscriptstyle H}-1|k)\big)$ such that constraint~\eqref{Eqn:con-velocity-all} is fulfilled and $\bar {\hat x}^{\scriptscriptstyle[N]}(k+\bar N_{\rm\scriptscriptstyle H}|k)\in \bar{\mathcal{X}}_{\rm\scriptscriptstyle F}^s$. 
Noting the fact that $\bar N_{\rm\scriptscriptstyle H}\geq N_{\alpha}$, one has $ \alpha(k+\bar N_{\rm\scriptscriptstyle H})=1\, \forall k\geq 0$. Let  $\overrightarrow {\Delta u_s^{\scriptscriptstyle[N],s}}(k+1:k+\bar N_{\rm\scriptscriptstyle H}+1|k+1)=\big(\Delta u_s^{\scriptscriptstyle[N],o}(k+1|k),\cdots,\break\Delta u_s^{\scriptscriptstyle[N],o}(k+\bar N_{\rm\scriptscriptstyle H}-1|k),\bar K_{s,\rm\scriptscriptstyle H}(\bar x^{\scriptscriptstyle[N]}(k+\bar N_{\rm\scriptscriptstyle H}|k)-\bar C^{\top}y_{s,r})\big)$ be the candidate input sequence at the next time instant $k+1$. 
As condition~\eqref{Eqn:state-terminal-size} is assumed, it holds that $\bar {\hat x}(k+\bar N_{\rm\scriptscriptstyle H}|k+1)\in \bar{\mathcal{X}}_{\rm\scriptscriptstyle F}$. In view of this, $\bar {\hat x}(k+\bar N_{\rm\scriptscriptstyle H}+1|k+1)\in \bar{\mathcal{X}}_{\rm\scriptscriptstyle F}^s$ can be verified in view of the definitions of $\bar K_{s,\rm\scriptscriptstyle H}$ and $\bar{\mathcal{X}}_{\rm\scriptscriptstyle F}^s$. Hence, the recursive feasibility of~\eqref{Eqn:HLoptimiz-im} follows.
}
\subsubsection{Convergence of the Incremental D-MPC}
In view of~\eqref{Eqn:alpha} and recalling the feasibility result of \eqref{Eqn:HLoptimiz-im} under $d$ being constant,  along the same line of Section~\ref{sec:theo-proof}, one can compute
\begin{equation}\label{Eqn:cost_monotonic1-im}
\begin{array}{l}
\bar J_{\rm\scriptscriptstyle H}^o(\bar x^{\scriptscriptstyle[N]}(k+1|k))-\bar J_{\rm\scriptscriptstyle H}^o(\bar x^{\scriptscriptstyle[N]}(k))\leq\\
-(\|\bar x_s^{\scriptscriptstyle[N]}(k)-\bar C^{\top} y_{s,r}\|^2_{Q_{s,\rm\scriptscriptstyle H}}+\|\Delta u_s^{\scriptscriptstyle[N],o}(k|k)\|^2_{R_{s,\rm\scriptscriptstyle H}})
\end{array}
\end{equation}
where $\bar J_{\rm\scriptscriptstyle H}^o$ is the optimal cost.
\eqref{Eqn:cost_monotonic1-im} implies that
$\bar J_{\rm\scriptscriptstyle H}^o(\bar x^{\scriptscriptstyle[N]}(k+1|k))-\bar J_{\rm\scriptscriptstyle H}^o(\bar x^{\scriptscriptstyle[N]}(k))$ converges to zero. Consequently, it holds that $\|\bar x_s^{\scriptscriptstyle[N]}(k)-\bar C^{\top} y_{s,r}\|^2_{\bar Q_{\rm\scriptscriptstyle H}}+\|\Delta u_s^{\scriptscriptstyle[N],o}(k|k)\|^2_{\bar R_{\rm\scriptscriptstyle H}}\rightarrow 0$ as well.  Recalling the definitions of $\bar Q_{\rm\scriptscriptstyle H}$ and $\bar R_{\rm\scriptscriptstyle H}$,  it holds that $\lim_{k\to+\infty} \bar x_s^{\scriptscriptstyle[N]}(k)=\bar C^{\top} y_{s,r}$  and $\lim_{k\to+\infty} \Delta u_s^{\scriptscriptstyle[N]}(k)=0$. Consequently, one has $\lim_{k\to+\infty} y^{\scriptscriptstyle[N]}(k)=y_r$, $\lim_{k\to+\infty} u_s^{\scriptscriptstyle[N]}(k)=const$. In view of Proposition~\ref{prop:obser-con}, it promptly follows that, $\lim_{k\to+\infty} x^{\scriptscriptstyle[N]}(k)=x_r$, $\lim_{k\to+\infty} u_s^{\scriptscriptstyle[N]}(k)=u_{s,r}$.
The arguments for the results $\lim_{h\to+\infty} \Delta u(h)=0$,
$\lim_{h\to+\infty} y(h)=y_r$ are similar to Section~\ref{sec:theo-proof}. Consequently, one has $\lim_{h\to+\infty} x(h)=x_r$, and $\lim_{h\to+\infty} u(h)=u_r$.
\hfill $\square$

 
%

\ifCLASSOPTIONcaptionsoff
  \newpage
\fi
 
\bibliographystyle{IEEEtran}
\bibliography{IEEEabrv,ref}
\end{document}